\documentclass[10pt]{article}

\usepackage[a4paper, left=2.5cm, right=2.5cm, top=3cm, bottom=3cm]{geometry}
\usepackage[authoryear]{natbib}
\usepackage{amssymb}
\usepackage{amsmath}
\usepackage{graphicx}
\usepackage{siunitx}
\usepackage{subfigure}
\usepackage{tikz} 
\usepackage{multirow}
\usepackage{booktabs}
\usepackage{float}
\usepackage{placeins}
\usepackage{booktabs} 
\usepackage{bm} 
\usepackage{adjustbox} 
\usepackage[labelfont=bf,labelsep=newline]{caption}

\title{Copula-based spatio-temporal modeling of air pollutant data incorporating covariate dependencies} 

\author{
 Soyun Jeon\\
  \small Department of Applied Statistics, Hanyang University, South Korea\\
  \and
 Jungsoon Choi\\
  \small Department of Mathematics, Hanyang University, South Korea \\ 
 \small Research Institute for Natural Sciences, Hanyang University, South Korea\\
 \small \texttt{jungsoonchoi@hanyang.ac.kr}
}

\date{}

\begin{document}

\maketitle

\let\thefootnote\relax
\footnotetext{This study is an extended version of master \textquotesingle s thesis of Soyun Jeon.} 

\begin{abstract}

Elevated levels of $\text{PM}_{10}$ are known to cause severe respiratory and cardiovascular diseases, and, in extreme cases, cancer and mortality. Despite various reduction policies implemented across different sectors, $\text{PM}_{10}$ concentrations in South Korea continue to exceed the annual recommended limits set by the World Health Organization. Spatio-temporal $\text{PM}_{10}$ concentrations may exhibit both spatial and temporal dependencies. Additionally, interactions between $\text{PM}_{10}$ and environmental factors can further influence the variability in $\text{PM}_{10}$. Therefore, this study proposes a method that incorporates the spatio-temporal neighbors of covariates alongside those of $\text{PM}_{10}$ by adopting an approach that explains spatio-temporal interactions through spatio-temporal neighbors. Vine copulas are used to integrate the pairwise dependence structures between a given location and its surrounding spatio-temporal neighbors. We applied the model to weekly average $\text{PM}_{10}$ data from South Korea in 2019, using $\text{PM}_{2.5}$ and $\text{CO}$ as covariates. Given that all three variables exhibited skewness, we assumed the Gumbel and Generalized Extreme Value distributions as marginal distributions. The proposed model outperformed a traditional Bayesian spatio-temporal model, a kriging method, and an alternative copula-based approach, particularly in predicting the top 5\% of extreme values, by effectively capturing tail dependencies crucial for extreme value analysis. This study highlights the importance of utilizing vine copulas to effectively model diverse dependency structures in spatio-temporal data while simultaneously accommodating spatial and temporal dimensions, including spatio-temporal dependencies among covariates. The results underscore the broader applicability of the proposed approach to other fields where complex dependency structures are present.
\end{abstract} 

\section{Introduction}
\label{sec1}
According to the Clean Air Conservation Act, particulate matter refers to tiny airborne particles, of which particles with a diameter of less than 10 $\mu$m are known as $\text{PM}_{10}$. The primary sources of $\text{PM}_{10}$ include minerals, road and soil dust, biomass combustion, and vehicle emissions, along with industrial and agricultural activities and natural disasters. Furthermore, sulfate, nitrate, and sodium generated by atmospheric chemical reactions also contribute to the formation of $\text{PM}_{10}$ \citep{kim2015review, mukherjee2017world}. Although exposure to $\text{PM}_{10}$ is primarily associated with respiratory diseases \citep{faustini2013air, liu2019ambient, feng2019short}, this particulate matter has also been linked to a variety of adverse health outcomes, including preterm birth, low birth weight, infant mortality, increased risk of premature death, asthma, diabetes, cancer, and cardiovascular diseases \citep{kim2015review, mukherjee2017world}. Recognizing its harmful health impacts, the International Agency for Research on Cancer (IARC) classified particulate matter as a group 1 carcinogen in 2013 \citep{iarc2013}. These findings underscore the significant and multifaceted threat that $\text{PM}_{10}$ poses to public health.

To mitigate the public health risks associated with exposure to $\text{PM}_{10}$, the World Health Organization (WHO) has established guideline limits of 15 $\mu\text{g}/\text{m}^3$ for annual average $\text{PM}_{10}$ concentrations \citep{world2021global}. However, recent data from the Korea Meteorological Administration have indicated persistent exceedance of these recommended limits in South Korea, with annual averages recorded at 36 $\mu\text{g}/\text{m}^3$ in 2021, 31 $\mu\text{g}/\text{m}^3$ in 2022, and 42 $\mu\text{g}/\text{m}^3$ in 2023. Given these persistent exceedances, providing reliable predictions of elevated $\text{PM}_{10}$ concentrations is crucial to facilitate effective and timely public health interventions and reduce associated health risks. Reliable forecasts of high $\text{PM}_{10}$ concentrations can enable proactive measures, such as issuing early alerts to encourage individuals to wear protective masks or limit outdoor activities. Moreover, administrative authorities can implement regulatory actions, including restricting vehicular traffic or reducing industrial emissions on days with predicted elevated pollution levels \citep{taheri2016statistical}. While such responses may not always result in immediate reductions in $\text{PM}_{10}$ concentrations, they can mitigate exposure and minimize public inconvenience by addressing harmful pollutants in advance. Thus, the development of an effective and accurate $\text{PM}_{10}$ prediction system is essential to protect public health and guide evidence-based interventions.

Research on predicting $\text{PM}_{10}$ concentrations has included temporal analysis \citep{gocheva2014time, cujia2019forecast,cekim2020forecasting}. Alternatively, spatial methods provide an approach to capture variation in $\text{PM}_{10}$ concentrations across different regions \citep{hamm2015spatially,aguirre2017developing,pradabmook2021estimation}. Among these methods, kriging is one of the most widely used spatial interpolation techniques, which is a linear estimation method using variogram-based weights to describe spatial correlation \citep{isaaks1989applied}. Many studies have applied kriging to spatial or spatio-temporal modeling of air pollutants such as $\text{PM}_{10}$ data \citep{sajjadi2017measurement,ahmed2018spatiotemporal,lin2018estimates,van2020spatio,shao2020estimating,rahman2023high,morillo2024spatial}. Bayesian approaches offer another option for modeling through spatio-temporal frameworks \citep{riccio2006hierarchical,cocchi2007hierarchical,cameletti2011comparing, pirani2014bayesian,mahdi2021hierarchical}. However, kriging faces challenges such as high computational cost as data size increases and the restrictive assumptions of Gaussian processes \citep{liao2006gis, gentile2013interpolating, cressie2015statistics}. These challenges are particularly relevant for $\text{PM}_{10}$ data, which often show skewness and heavy tails \citep{lee2006study}, deviating significantly from normality. Similarly, Bayesian inference via Markov Chain Monte Carlo (MCMC) methods is computationally demanding and encounters convergence issues, especially when applied to large-scale spatio-temporal datasets. Given these limitations, the need for alternative methodologies that balance computational efficiency with the ability to model high-dimensional and asymmetric spatio-temporal $\text{PM}_{10}$ data is growing.

Copulas are powerful tools for modeling dependence structures between variables, particularly to capture asymmetric, non-linear, and skewed relationships \citep{patton2004out, gregoire2008using, haslauer2010application, zhang2018modeling}. These strengths make copula-based models highly effective for non-Gaussian data and often superior to kriging methods in such contexts. For instance, \citet{bardossy2008geostatistical} and \citet{agarwal2021copula} illustrated the efficacy of copula-based approaches to handle skewed and non-Gaussian data. \citet{musafer2013copula} and \citet{gnann2018improving} highlighted the ability of copulas to model nonlinear dependencies that kriging may overlook. Copula-based models are also highly effective in capturing tail dependencies, which makes them particularly well-suited for extreme value analysis in environmental datasets \citep{salvadori2007extremes, davison2012statistical}. For example, \citet{chan2018multivariate} proposed an innovative copula-based spatial extreme value model to characterize the spatial dependence of several extreme air pollution levels. \citet{stein2020spatial} applied copula models to estimate extreme concentrations in unmonitored regions, focusing on $\text{PM}_{1}$ (particulate matter with a diameter of less than 1 $\mu$m). Similarly, \citet{fuentes2013nonparametric},\ \citet{durocher2016prediction}, and \citet{yu2016modeling} provided substantial evidence supporting the effectiveness of copula-based approaches in modeling spatial extreme dependencies in environmental datasets. These features highlighted the practical advantages of copula-based models in tackling the challenges associated with non-Gaussian and spatial extreme value data.

Copulas can be extended to multivariate settings, but their flexibility diminishes, and structural complexity increases as the dimensionality of the data increases \citep{brechmann2013modeling}. To address these challenges, vine copulas have emerged as an alternative framework to model high-dimensional multivariate data \citep{joe2011dependence, brechmann2013modeling, omidi2022spatial,ismail2023modeling}. As a class of multivariate copulas, vine copulas decompose complex multivariate distributions into a series of bivariate copulas, thereby enabling the flexible representation of intricate dependence structures without imposing restrictive assumptions. Effectiveness in handling multivariate data had been demonstrated across various domains in spatial analysis. \citet{bardossy2011interpolation} applied Gaussian copulas and non-Gaussian vine copulas for spatial interpolation of multiple groundwater quality parameters, showing that these copulas outperformed kriging in interpolation accuracy. Moreover, vine copula-based approaches provided more realistic confidence intervals compared to their Gaussian alternatives. \citet{khan2020novel} compared the abilities of vine copulas and Student\textquotesingle s t copulas to capture multivariate spatial dependence structures in climate variables such as temperature and precipitation, finding that vine copulas generally exhibited superior performance. \citet{alidoost2021multivariate} developed a multivariate copula-based quantile regression model using vine copulas and applied it to spatial interpolation of daily temperature data in regions with missing observations, demonstrating its effectiveness in bias correction. Similarly, the application of vine copula-based quantile regression proved useful in modeling spatial and spatio-temporal patterns of COVID-19 in Italy\citep{d2022d}. Beyond environmental data, vine copulas have proven to be highly applicable in other high-dimensional settings, such as wind dynamics, medical research, reliability analysis, and financial modeling \citep{allen2013financial,jiang2015vine, erhardt2015spatial, montes2016multivariate,musafer2017non, khuntia2019multivariate, tosunoglu2020multivariate,nazeri2022multivariate}. These studies suggested that vine copulas can be flexible and powerful tools for modeling complex dependence structures in multivariate and high-dimensional contexts.

\citet{graler2014developing} introduced a spatio-temporal covariate vine (STCV) copula model that focused on the spatial and temporal neighbors of the dependent variable. By additionally incorporating the dependency between the dependent variable and a single covariate co-located in both space and time with this variable, he demonstrated the improved predictive performance compared to models without covariates. Furthermore, this approach provided a framework for spatio-temporal interpolation suitable for non-Gaussian data. Based on this study, \citet{alidoost2018copula} proposed a method that applied conditional probabilities based on multiple observed covariates from the nearest neighbors. Their approach showed superior predictive performance over models using a single covariate, however their analysis was confined to spatial data. In contrast, in the context of $\text{PM}_{10}$ space-time data, interactions with multiple covariates are likely to exhibit complex spatial and temporal dependencies \citep{chu2015modeling,xie2015spatiotemporal}. Incorporating these interactions into the model is expected to improve both estimation accuracy and predictive performance. 

Spatio-temporal variation presents significant challenges in capturing the complex dependency structures across spatial and temporal domains, particularly as the scale of the study area expands. While previous studies have demonstrated the effectiveness of vine copulas in addressing such challenges, to our knowledge, their application to spatio-temporal modeling in South Korea, specifically in the context of $\text{PM}_{10}$ concentrations, has not been explored. Therefore this study employs a vine copula-based approach to model the intricate spatio-temporal dependencies in $\text{PM}_{10}$ data. Building on the framework proposed by \citet{graler2014developing}, we propose an approach that not only captures the spatio-temporal variability of $\text{PM}_{10}$ but also simultaneously accounts for its interactions with two covariates across both space and time. These interactions are represented through the construction of spatio-temporal neighborhoods. By explicitly incorporating such dependencies, our approach refines the existing framework, offering a more comprehensive understanding of $\text{PM}_{10}$ dynamics in relation to the selected covariates. This study proves the effectiveness of the proposed model's spatio-temporal interpolation method for skewed data, in comparison with kriging and a Bayesian spatio-temporal model, which are widely used in spatio-temporal analysis. This approach aims to improve the predictive accuracy of $\text{PM}_{10}$ concentrations by accounting for additional dependencies. Furthermore, recognizing the ability of copulas to model tail dependencies \citep{joe2010tail}, this study seeks to explore ways to enhance predictive performance in extreme value data. This contribution is expected to provide valuable insights for policymakers to mitigate the adverse public health impacts of exposure to $\text{PM}_{10}$.

The study consists of five chapters. Chapter 1 provides an overview of the research background, previous research, and objectives of the study. Chapter 2 introduces the methodology used in the study. Chapter 3 provides a description of the actual data, and Chapter 4 presents the results derived from analysis using this data. Finally, Chapter 5 contains the conclusions of this study and suggestions for future studies.

\section{Methodology}
\label{sec2}
In Section 2.1, an overview of copula functions and their connection to joint probability distributions is provided. STCV model is introduced in Section 2.2, along with a discussion on parameter estimation using Kendall\textquotesingle s tau. We build upon the approach in Section 2.2 by incorporating spatio-temporal neighbors of covariates in Section 2.3 to better capture dependencies between covariates and the dependent variable. Bayesian spatial models used for comparison are briefly described in Section 2.4, and the performance metrics applied for evaluation are outlined in Section 2.5.

\subsection{Copulas}
\label{subsec2.1}
The copulas, initially introduced by \citet{sklar1959fonctions}, are multivariate cumulative distribution functions (CDF) of random variables\ $U_{1},\cdots,U_{n}$\ respectively satisfying a uniform distribution on the closed interval [0,1]. According to Sklar\textquotesingle s theorem, for random variables $X_{1},\cdots,X_{n}$ with a joint CDF denoted by $F$ and a set of marginal CDFs for each variable denoted by $F_1,\cdots,F_n$, there exists a copula function $C$ such that $F(x_{1},\cdots,x_{n})=C(F_1(x_{1}),\cdots,F_n(x_{n}))$.  Given $F_i(x_{i})=u_{i}$ for $u_i\in [0,1]$ and each $F_i^{-1}$ denotes the inverse function of $F_{i}$ for $i \in \{1,\cdots,n\}$, the lemma of Sklar\textquotesingle s theorem is the copula function $C$ that satisfies $C(u_{1},\cdots,u_{n})=F(F^{-1}_1(u_1),\cdots,F^{-1}_n(u_n))$. By the above theorem, the joint probability density function (PDF) of dependent variables can be decomposed, as in Equation (1), into the copula function and the marginal PDFs of the variables. Here, the copula function captures the dependency structure among the variables. For continuous random variables $X_{1}, \cdots,X_{n}$, this decomposition of the multivariate distribution into marginal distributions and the copula function is unique and corresponds mathematically to deriving the joint PDF \citep{sklar1959fonctions,nelsen2006introduction}.
\begin{equation} 
f(x_{1},\cdots,x_{n}) = c(u_{1},\cdots, u_{n}){\displaystyle \prod_{i=1}^{n} f_{i}(x_{i})} , \ 
\text{where} \  c(u_{1},\cdots, u_{n})=\frac{\partial^{n} C(u_{1},\cdots,u_{n})}{\partial u_{1}\times \cdots \times \partial u_{n}}.
\end{equation}

\subsection{STCV}
\label{subsec2.2}
\citet{graler2014developing} introduced a bivariate spatial copula $c_{h}$ that modeled spatial dependence as a function of the separation distance. As shown in Equation (2), this copula was expressed as a weighted linear combination of spatial copulas $c_{b,h}$ defined for each spatial bin $b$. The weights, $\lambda_{b} = \frac{h-l_{b-1}}{l_{b}-l_{b-1}}$ were computed via linear interpolation, where $h$ represented the separation distance, and $l_{1}, \cdots, l_{b}$ denoted the mean distances within each bin. This framework was extended to incorporate temporal dependence by representing the bivariate spatio-temporal copula at time $t$ as $c^{t}_{h}$ for $t=1, \cdots, T$ using a linear combination of distinct bivariate spatial copulas across different time points, where $T$ denotes the total number of time points.
\begin{equation}
c_{h} =
\begin{cases} 
c_{1,h} & \!   0 \!   \leq h  \! <l_{1} \\ 
         (1- \!  \lambda_{2}) \! \times  \!c_{1,h} \! + \! \lambda_{2} \! \times c_{2,h} &   \! l_{1}  \! \leq h  \!  <l_{2} \\ 
     \  \  \ \  \ \ \ \ \ \ \ \ \    \  \ \ \ \ \ \ \  \vdots \\
         (1- \! \lambda_{b}) \! \times c_{b-1,h} \! + \lambda_{b} \!\times \! 1 & \!  l_{b-1} \!  \leq h  \! <l_{b} \\
         1  & \! l_{b} \! \leq h
\end{cases}
\end{equation}

To estimate the parameters of a bivariate spatio-temporal copula, the data were partitioned into spatial bins at each time point. Kendall\textquotesingle s tau, which measures dependence independently of marginal distributions, enables direct modeling of dependence at the copula level \citep{graler2014developing}. For each time point, a correlogram based on Kendall\textquotesingle s tau was constructed to examine the dependence structure of the binned data. In copula families where a one-to-one relationship exists between Kendall\textquotesingle s tau and the copula parameter, polynomial functions of distance were fitted to the correlogram. These estimated polynomials were then used to determine the distance parameter as the sole parameter of the bivariate spatial copula, and allowed the copula to accurately reproduce Kendall\textquotesingle s tau at all distances. Using this parameter estimation approach, several copula families exhibiting a one-to-one correspondence with Kendall\textquotesingle s tau were fitted, and the family that maximized the log-likelihood within each spatial bin was selected. In this framework, the copula family depended only on distance, whereas the copula parameter varied with both distance and time.

To model spatio-temporal dependencies of the dependent variable, each observation of the dependent variable was treated as a central site, around which a spatio-temporal neighborhood was defined. Within this neighborhood, \citet{graler2014developing} modeled pairwise dependencies between locations over time using bivariate spatio-temporal copulas. Since these bivariate copulas model pairwise dependencies among neighbors but do not capture their joint dependence structure, vine copulas were used to integrate them into a single $n$-dimensional multivariate distribution. Vine copulas are graphical models composed of a hierarchical tree structure with multiple trees, where each tree represents dependencies between variables using bivariate copulas as fundamental building blocks. The first tree consists of $n-1$ bivariate copulas that account for all possible pairs of the dependent variables. Subsequent trees incorporate additional dependencies by conditioning on the distributions defined in previous trees, introducing new bivariate copulas at each stage. As a result, a total of $\frac{n(n-1)}{2}$ bivariate copulas are required to characterize the full $n$-dimensional dependence structure. By decomposing high-dimensional dependencies into unrestricted bivariate components, vine copulas offer remarkable flexibility in modeling complex dependence structures. They are particularly effective in capturing asymmetric and tail dependencies while allowing conditional independence in specific parts of the dependence structure \citep{joe1997multivariate, joe2010tail, erhardt2015spatial, okhrin2017copulae}. However, the model was limited to capturing dependencies within the spatio-temporal neighborhood of dependent variables.

\subsection{Proposed model : STCV with spatio-temporal neighbors of two covariates}
\label{subsec2.3}

Motivated by the limitations outlined in Section 2.2, we propose a comprehensive approach that simultaneously accounts for the spatio-temporal neighborhoods of both the dependent variable and two covariates. Let $d$ and $d_{c}$ represent the dimensions of the spatio-temporal neighborhoods for the dependent variable and two covariates, $v$ and $w$, respectively. For each observation of the dependent variable, a dataset of spatio-temporal neighbors is constructed, consisting of $d$ neighbors of the dependent variable and two distinct sets of $d_{c}$ covariate neighbors, which results in a total of $d^{\prime} = d + 2d_{c}$ neighboring values. By applying the CDF of the marginal distributions to transform this dataset into the unit interval $[0,1]$, the transformed values are defined as the central site ($u_{0}$), the spatio-temporal neighbors of the dependent variable ($u_{i} \  \text{for} \ i=1,\cdots,d$), and the spatio-temporal neighbors of the covariates ($u_{v_{j}}$ and $u_{w_{j}}$, for $j=1,\cdots,d_{c}$).

The proposed model constructs the first tree using three bivariate spatio-temporal copulas, each capturing a specific dependency structure within the defined neighborhoods. The first copula, $c_{0,i}^{t}$, characterizes the dependence between $u_{0}$ and $u_{i}$. The second and third copulas, $c^{t}_{v_{0,j}}$ and $c^{t}_{w_{0,j}}$, describe the dependence between $u_{0}$ and $u_{v_{j}}$, and between $u_{0}$ and $u_{w_{j}}$, respectively. These copulas collectively capture both the spatio-temporal dependence of the dependent variable and covariate-driven dependencies, forming the foundation of the vine copula construction. To effectively capture the dependencies between a central location and its associated spatio-temporal neighbors, a C-vine structure is employed, where the central node serves as the primary connecting point across all trees. For notational simplicity, let $v_{k}^{-l}=(v_{1},\cdots,v_{k-l})$ and $w_{k}^{-l}=(w_{1},\cdots,w_{k-l})$ for $l=1,2$. Given $d^{\prime}$, for $1 \leq k < d^{\prime}$ and $0 \leq m \leq d^{\prime}-k$, each subsequent tree is constructed based on values determined in the preceding tree. These relationships are formally described in Equation (3), which captures the conditional dependencies across trees.
\begin{align}
 & u_{k+m|v^{-1}_{k},w^{-1}_{k},0,1,\cdots,k-1}   \nonumber\\
 & = F_{k+m|v^{-1}_{k},w^{-1}_{k},0,1,\cdots,k-1}(u_{k+m}| u_{v_{1}},\cdots,u_{v_{k-1}}, u_{w_{1}},\cdots,u_{w_{k-1}},u_{0},u_{1},\cdots,u_{k-1}) \\
& = \frac{\partial C_{k-1,k+m|v^{-2}_{k},w^{-2}_{k},0,1,\cdots,k-2}(u_{k-1|v^{-2}_{k},w^{-2}_{k},0,1,\cdots,k-2},u_{k+m|v^{-2}_{k},w^{-2}_{k},0,1,\cdots,k-2})}{\partial u_{k-1|v^{-2}_{k},w^{-2}_{k},0,1,\cdots,k-2}}. \nonumber
\end{align}

\noindent The joint density of the proposed model is expressed as the product of all relevant bivariate copula densities, as shown in Equation (4).
\begin{align}
&c_{h}^{t}(u_{v_{1}},\cdots,u_{v_{d_{c}}}, u_{w_{1}},\cdots,u_{w_{d_{c}}},u_{0},u_{1},\cdots,u_{d}) \nonumber\\
& = \displaystyle \prod_{j=1}^{d_{c}} c^{t}_{v_{0,j}}(u_{0},u_{v_{j}})\times \displaystyle \prod_{j=1}^{d_{c}} c^{t}_{w_{0,j}}(u_{0},u_{w_{j}})\times  \displaystyle \prod_{i=1}^{d} c_{0,i}^{t}(u_{0},u_{i})  \\
&\times \displaystyle \prod_{k=1}^{d^{\prime}-1}\prod_{m=1}^{d^{\prime}-k} c_{k,k+m|v^{-1}_{k},w^{-1}_{k},0,\cdots,k-1}(u_{k|v^{-1}_{k},w^{-1}_{k},0,\cdots,k-1},u_{k+m|v^{-1}_{k},w^{-1}_{k},0,\cdots,k-1}).\nonumber
\end{align}

The parameter estimation for the proposed model begins with the first tree, which consists of bivariate spatio-temporal copulas and can be estimated as described in Section 2.2. For subsequent trees, conditional spatio-temporal data are constructed using the copulas estimated from the previous tree, as detailed in Equation (3). The copulas that maximize the log-likelihood are sequentially estimated.

To predict the dependent variable at unobserved locations, we utilize the $d^{\prime}$ nearest spatio-temporal neighbors, denoted as $\tilde{u_{1}}, \cdots, \tilde{u_{d^{\prime}}}$. The predictive mean $\hat{Y}_{\text{mean}}(s_{\text{pred}})$ and quantile estimate $\hat{Y}_{\text{p}}(s_{\text{pred}})$ at the target location $s_{\text{pred}}$ are defined by Equation (5). Here, $F_{\text{pred}}^{-1}$ represents the inverse CDF for the marginal distribution at $s_{\text{pred}}$, and $c$ denotes the proposed model associated with $s_{\text{pred}}$, while $C$ refers to the CDF of $c$. This approach illustrates that accurate predictions can be made without requiring the complete joint PDF, provided the copula distribution is known. Furthermore, as the method relies on the probabilistic integral transform, it facilitates realistic uncertainty quantification \citep{graler2014developing}.
\begin{align}
\hat{Y}_{\text{mean}}(s_{\text{pred}}) 
                            & = \displaystyle \int_{[0,1]} F_{\text{pred}}^{-1} \times c(u|\tilde{u_{1}}, \cdots, \tilde{u_{d^{\prime}}}) du.  \nonumber \\
  \hat{Y}_{\text{p}}(s_{\text{pred}}) & = F_{\text{pred}}^{-1} \times {C}^{-1}(p|\tilde{u_{1}}, \cdots, \tilde{u_{d^{\prime}}}).
\end{align}

\subsection{Bayesian spatio-temporal model}
\label{subsec2.4}

We evaluate the effectiveness of copula modeling by comparing it with a traditional Bayesian spatio-temporal model such as Equation (6) \citep{cocchi2007hierarchical},
\begin{equation} 
      y_{st} = \beta_{0}+\beta_{1}v_{st}+ \beta_{2}w_{st}+\gamma_{t}+\tau_{t}+v_{s}+u_{s}+e_{st},\ \text{where} \ s= 1,2,\cdots,S \ \text{and} \  t = 1,2,\cdots,T.
\end{equation}
Let $y_{st}$ represent the dependent variable at spatial location $s$ and time $t$, while $v_{it}$ and $w_{it}$ refer to covariates at the same location and time, where $S$ and $T$ are the number of spatial and temporal observations, respectively. The regression coefficient $\beta_0,\ \beta_1,\ \beta_2$ are considered as constants regardless of each space and time. Temporal independence is captured by the vector $\gamma_{t}$, where $\gamma_{t}\sim N(0,\epsilon_{\gamma}^2)$, while temporal dependence is modeled via an AR(1) process in the vector $\tau_{t}$, such that $\tau_{t}=\phi\tau_{t-1}+\varepsilon_{t}$, with $\varepsilon_{t} \sim N(0,\epsilon_{\varepsilon}^2)$ and $-1<\phi<1$. Similarly, spatial independence is represented by the vector $u_{s}$, with  $u_{s} \sim N(0,\epsilon_{u}^2)$, while spatial correlation is modeled by the vector $v_{s}$ using a spatial power exponential family function $f(d_{s_{i}s_{j}};\rho,\kappa)=\exp[-(\rho d_{s_{i}s_{j}})^{\kappa}]$, where $\rho > 0$ and $\kappa \in (0, 2]$. Here, $d_{s_{i}s_{j}}$ denotes the distance between regions $s_{i}$ and $s_{j}$, with $\rho$ representing the rate of rapid correlation decay with increasing distance and $\kappa$ controlling the smoothness of spatial variation. Finally, the error term $e_{st}$ is assumed to follow independent and identically distributed normal distribution, $e_{st} \sim N(0,\epsilon_{e}^2)$.

\subsection{Evaluation}
\label{subsec2.5}

The mean absolute error (MAE) measures the average magnitude of absolute prediction errors, providing a straightforward measure of overall model accuracy. In contrast, the root mean squared error (RMSE) calculates the square root of the mean squared prediction errors, which gives greater weight to larger errors and makes it more sensitive to outliers while emphasizing significant deviations in prediction. Since smaller differences between observed and predicted values indicate better model performance, models with lower MAE and RMSE are considered to have superior predictive accuracy. To enhance the reliability of model evaluation, a 10-fold cross-validation (CV) approach was used for both model fitting and prediction.

\section{Data}
\label{sec3}
In this study, we used the daily average data for $\text{PM}_{10}$, $\text{PM}_{2.5}$ (particulate matter with a diameter of 2.5 $\mu$m or less), and $\text{CO}$ in South Korea, obtained from the Korea Environment Corporation via the AirKorea platform (https://www.airkorea.or.kr/), with $\text{PM}_{10}$ as the dependent variable and $\text{PM}_{2.5}$ and $\text{CO}$ as the covariates. According to the Korea Environment Corporation, missing values may occur at different air quality monitoring stations due to test operations of monitoring devices, performance checks, regular inspections, and equipment calibration (e.g., zero and span calibrations). Based on these considerations, we limited our spatial domain to 409 out of 464 urban and suburban monitoring stations where the missing rates for both the dependent variable and the covariates were below 20\%. The remaining missing values were interpolated using ordinary kriging, with covariance models chosen from the spherical or exponential model at each time point based on the smallest sum of squares, with the binning set to 50 km. The daily average data from January 1 to December 29, 2019 were aggregated into weekly average data to mitigate daily variability. 

The summary statistics of the three variables, converted to weekly averages within the spatial and temporal domains analyzed, are presented in Table 1. All three variables exhibited skewed distributions with heavy right tails. To estimate the marginal distribution, it is necessary to select an appropriate model that accounts for the observed skewness. 
\FloatBarrier 
\begin{table}[H]
\centering
\caption{Summary statistics of the variables.}\label{table:table1}
\begin{tabular}{@{}cS[table-format=2.2] S[table-format=2.2] S[table-format=2.2] S[table-format=3.2] S[table-format=2.2]@{}}
\toprule
 \textbf{Variables} & \textbf{{Mean}} & \textbf{{Min}} & \textbf{{Median}} & \textbf{{Max}}  &  \textbf{{SD}} \\
\midrule
 \textbf{$\text{PM}_{10}(\mu\text{g}/\text{m}^3)$} & 41.62 &  7.00 & 39.00 & 136.00 & 18.14 \\
 \textbf{$\text{PM}_{2.5}(\mu\text{g}/\text{m}^3)$} & 23.45 &  2.00 & 21.00 & 94.00 & 12.63\\ 
 \textbf{$\text{CO(ppm)}$} & 0.48 &  0.01  & 0.46  & 1.48  & 0.16  \\
\bottomrule
\end{tabular}
\end{table}   

Figure 1 illustrates the time series of the monthly average $\text{PM}_{10}$ concentrations across all monitoring stations. The numerical labels, in blue text, indicate the corresponding months. The shaded region highlights the period from January to April, during which relatively high concentrations were observed. Figure 2 shows the monthly average $\text{PM}_{10}$ concentrations observed at each monitoring station from January to April. One of the regions that exhibited consistently high $\text{PM}_{10}$ concentrations was the Seoul Metropolitan Area (SMA), which includes Seoul (the capital of South Korea), Gyeonggi Province which surrounds Seoul), and Incheon (a major port city to the west of Seoul). For clarity, the SMA is marked with black dashed circles in Figure 2.

The time series plot and maps revealed the spatio-temporal correlation of $\text{PM}_{10}$ concentrations, underscoring the need for a spatio-temporal approach to data analysis. To better understand the elevated pollutant concentrations observed from January to April and in the SMA, we restructured the spatio-temporal domain into two datasets: D1 represented nationwide data for the entire year (409 locations over 52 weeks), and D2 was confined to the SMA for January to April (153 locations over 17 weeks).

\begin{figure}[t]
\centering
\includegraphics[width=\textwidth]{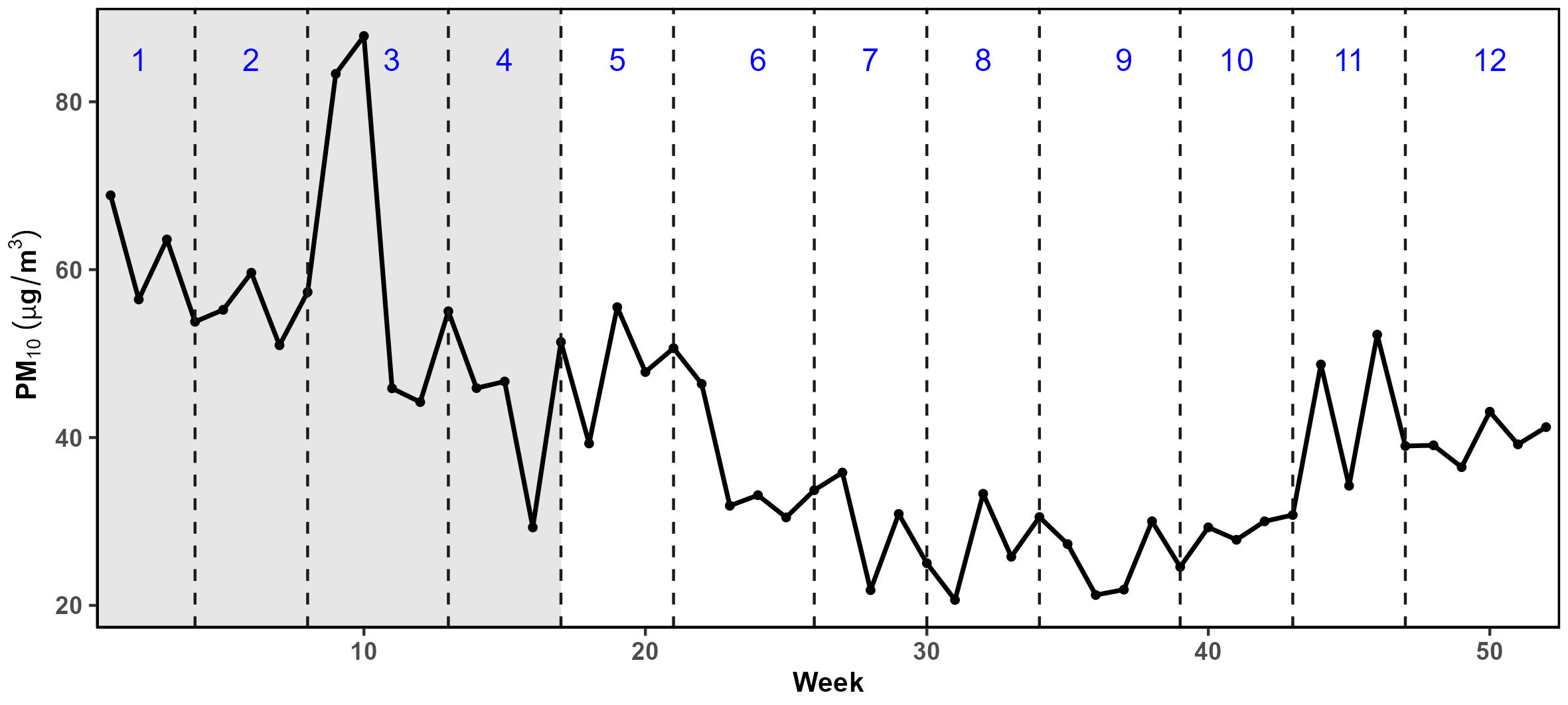} 
\caption{The time series of the $\text{PM}_{10}$ data.}\label{fig:fig 1}
\end{figure}

\section{Results}
\label{sec4}
The results section was organized into five subsections to provide a detailed overview of the analysis. Transformation of the data to conform to the uniform distribution on $[0,1]$ was conducted as a prerequisite for the application of copula functions, as detailed in Section 4.1. The estimation of the first tree in the proposed model was discussed in Section 4.2, with particular emphasis on the spatial binning process. The subsequent trees were estimated using spatio-temporal neighbor data, revealing a diversity of copula families in Section 4.3. The estimation results for the four comparative models were presented in Section 4.4. Predictive accuracy was then assessed in Section 4.5, considering overall predictions, extreme values, and the proportion of observed values within the 95\% prediction intervals.  To avoid redundancy, Figures 3, 4, and 5 use blue to represent results under the assumption of a Gumbel distribution and orange for those based on a generalized extreme value (GEV) distribution. Furthermore, Figures 4, 5, and 6 present the aggregated results in all the folds of the 10-fold CV, while Tables 2 and 3 summarize the outcomes as the average in all folds.

\subsection{Marginal distribution of the variables}
\label{subsec4.1}
Figure 3 presents histograms of each variable in the dataset D1, which revealed a left-skewed distribution for all variables. Based on these observations, the marginal distributions for each region were estimated using the Gumbel or generalized extreme value (GEV) distributions to effectively capture regional characteristics. The CDFs for these distributions are detailed in Appendix A for reference. Using the estimated regional parameters, the inverse CDF was applied to transform all variables into uniform distributions. The transformed results of each marginal distribution are represented by solid lines in Figure 3, which closely match the shapes of the histograms. Kolmogorov-Smirnov tests conducted at all observation sites confirmed the appropriateness of the transformation for the Gumbel and GEV distributions, validating that the transformation effectively preserved the data characteristics. A similar analysis applied to the dataset D2 yielded comparable results, further validating the robustness of this approach.

\subsection{Estimation of the first tree in the proposed model: bivariate spatio-temporal copulas of the dependent variable and two covariates}
\label{subsec4.2}
The number of spatial bins requires careful consideration since too many bins may result in insufficient observations per bin, whereas too few could constrain the model's flexibility. Based on the total size of the data and information from previous studies, the number of spatial bins was set to 10 to ensure adequate data within each bin while maintaining the model flexibility. Moreover, we considered contemporaneous (lag 0) and one-time lag, since Kendall\textquotesingle s tau for spatial bins was significantly non-zero at these lags but zero at longer lags. The autocorrelation function (ACF) analysis further supported their inclusion.

Figure 4 presents the correlogram of fitted Kendall\textquotesingle s tau values across all spatial bins, with solid lines representing correlations at the same time point and dashed lines indicating correlations with a one-time lag. In the dataset D1, the correlation ranged between 0.38 and 0.72, whereas in the dataset D2, it ranged from 0.20 to 0.60. For both datasets, the correlation at the same time point decreased as the spatial distance increased. In contrast, correlations with a one-time lag were negligible in relation to distance but remained slightly lower than those observed at the same time point.

The parameters for the candidate bivariate spatio-temporal copulas—Gaussian, Student\textquotesingle s t, Clayton, Frank, Gumbel, and Joe—were estimated using the polynomial in Figure 4. Figure 4 presents a summary of the 10-fold CV results for the selected bivariate spatio-temporal copulas that exhibited the highest log-likelihood values with different marginal distributions for each bin. Regardless of the assumed marginal distribution or data type, the T-copula and Gumbel copula indicating tail dependence were frequently selected at the same time point. Under a one-time lag condition, Gaussian or Frank copulas, which did not exhibit tail dependence, were predominantly chosen.

\FloatBarrier 
\begin{figure}[t]
\subfigure[\fontsize{10}{15}\selectfont January]{{\includegraphics[width=0.5\textwidth]{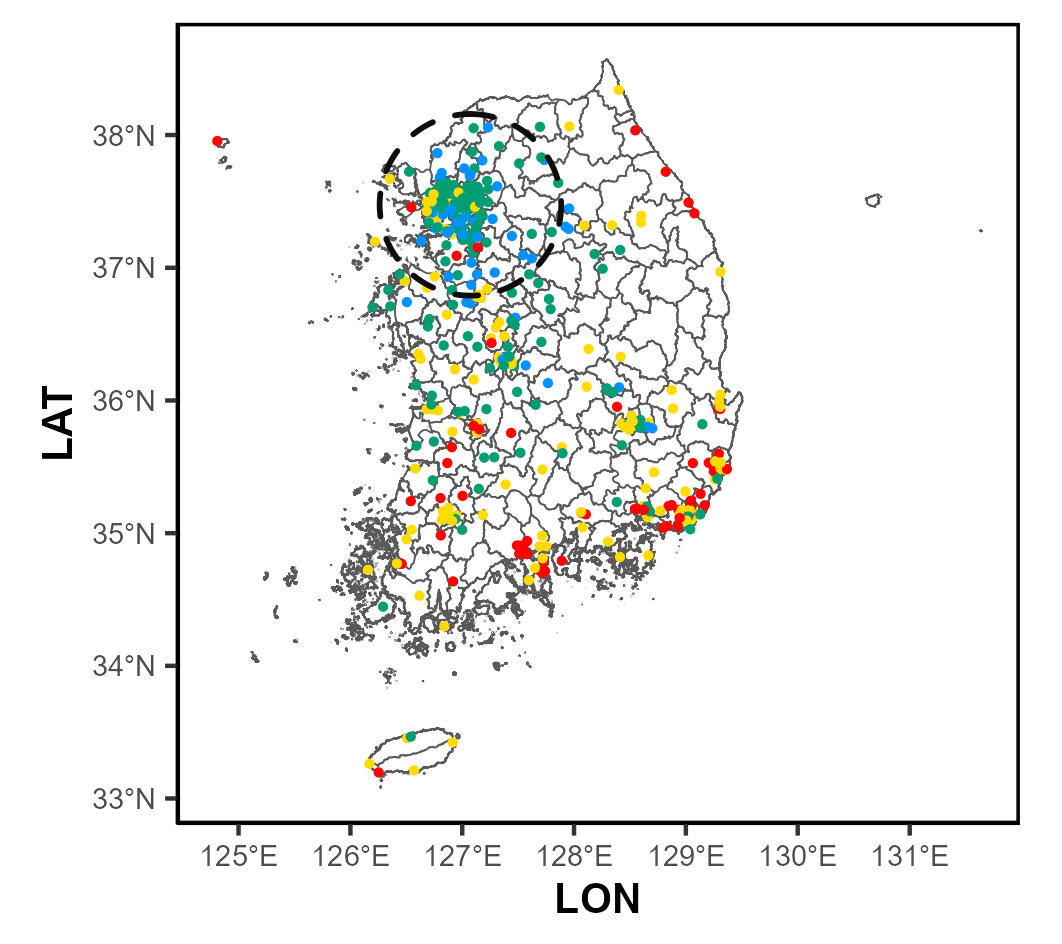} }}%
\subfigure[\fontsize{10}{15}\selectfont February]{{\includegraphics[width=0.5\textwidth ]{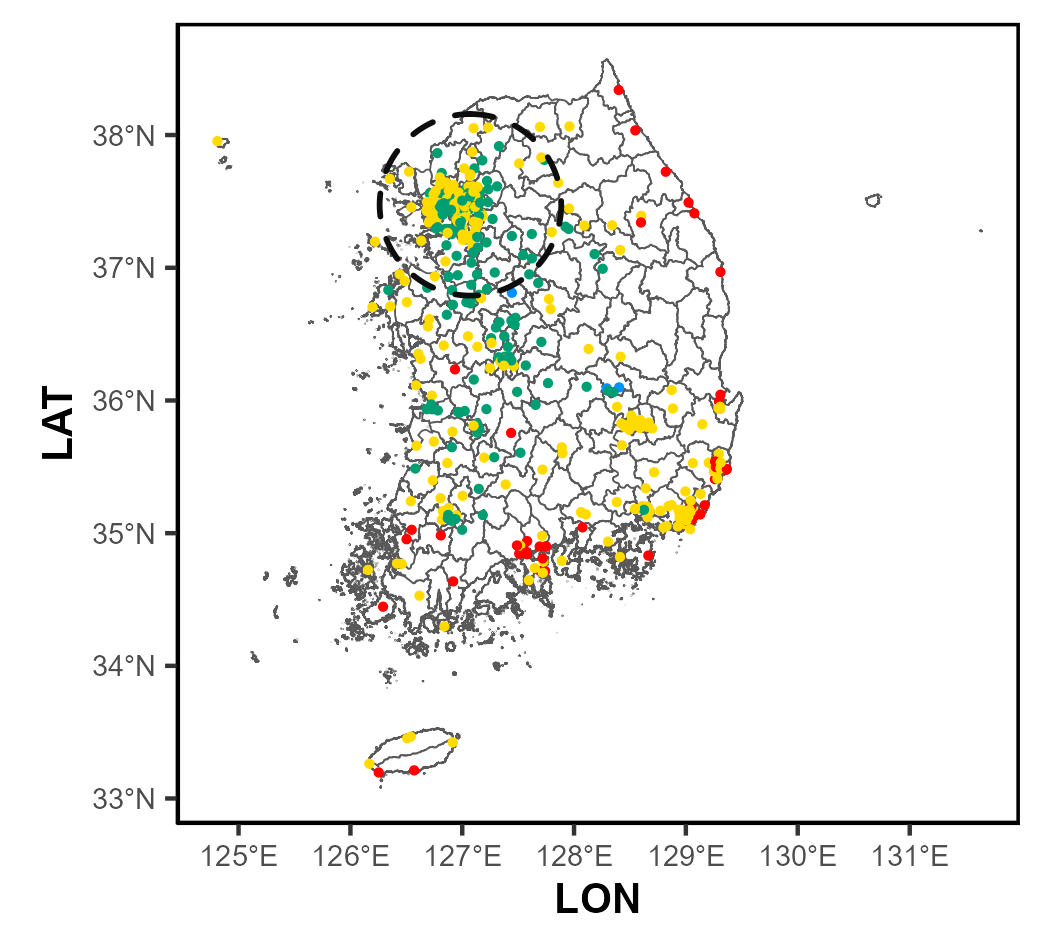} }}%
\hfill 
\subfigure[\fontsize{10}{15}\selectfont March]{{\includegraphics[width=0.5\textwidth ]{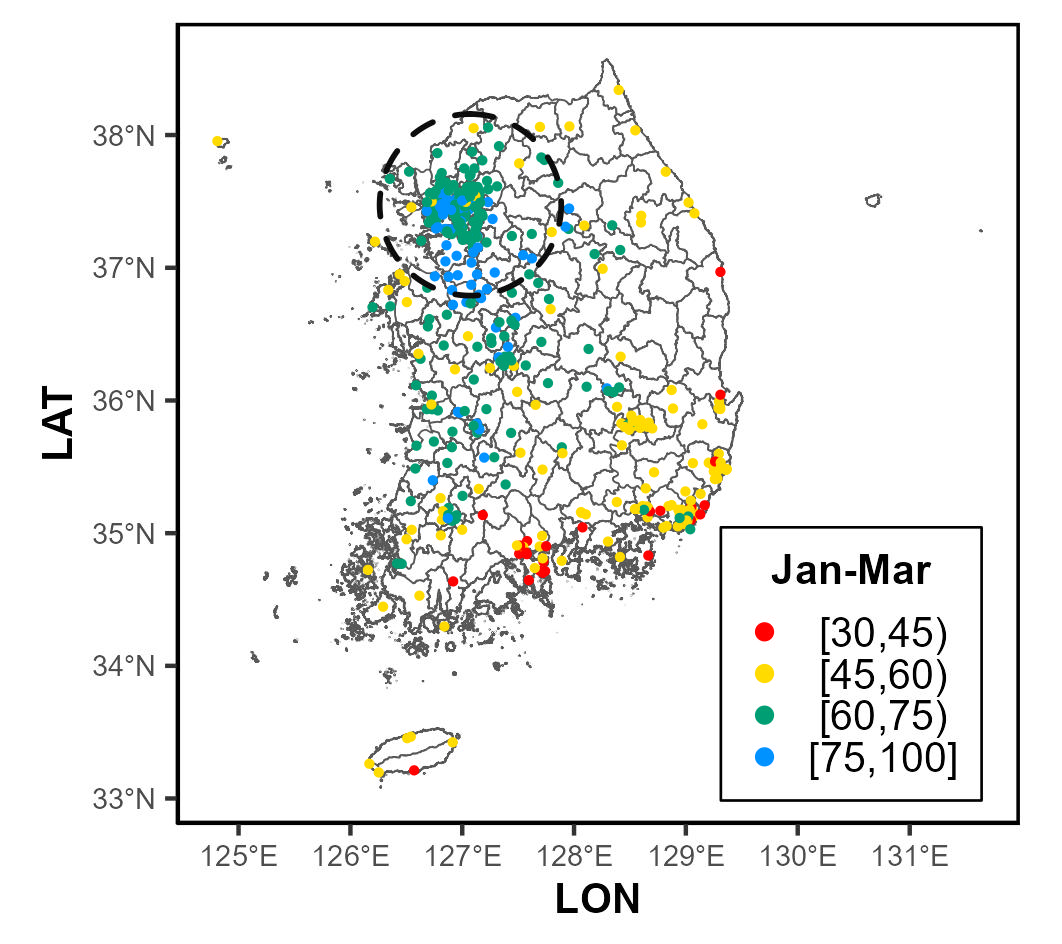} }}%
\subfigure[\fontsize{10}{15}\selectfont April]{{\includegraphics[width=0.5\textwidth ]{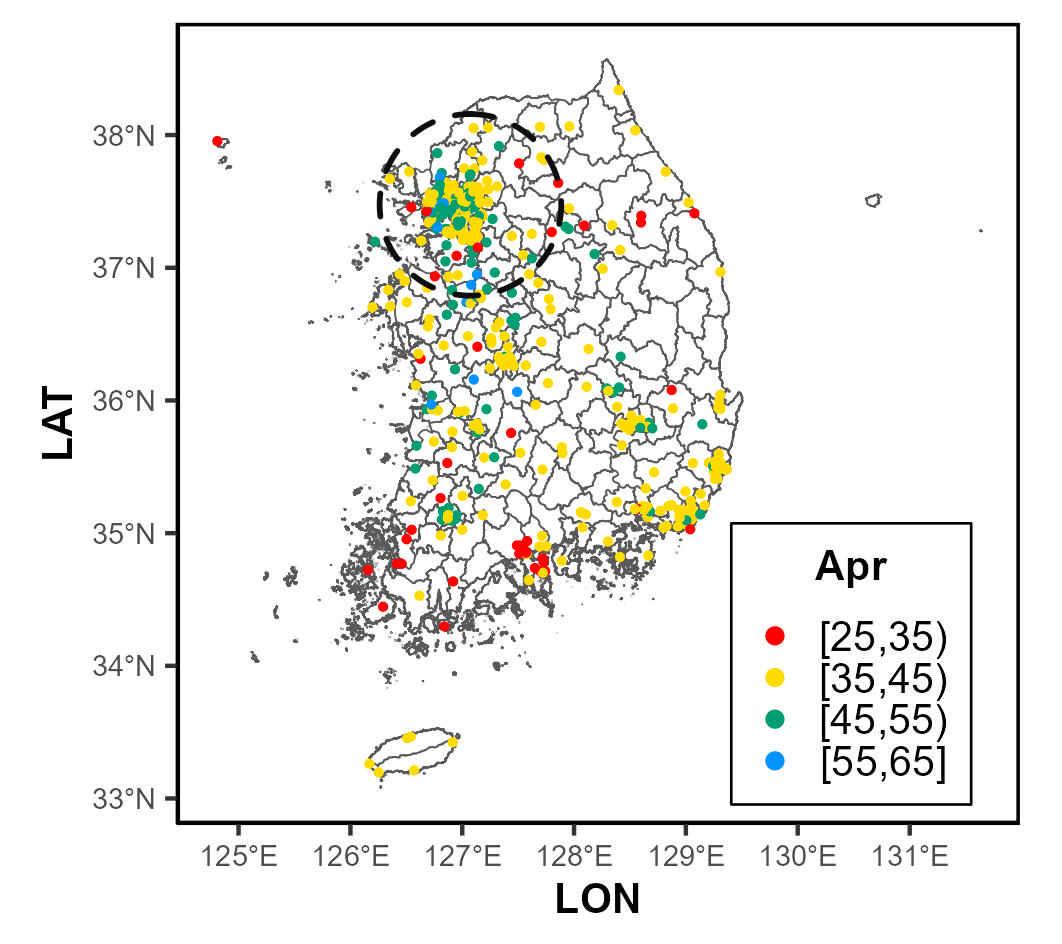} }}%
\caption{Maps of the monthly average $\text{PM}_{10}$ concentrations each month (January - April).}\label{fig:fig 2}
\end{figure}

\FloatBarrier 
\begin{figure}[h]
\centering
\subfigure[\fontsize{10}{15}\selectfont $\text{PM}_{10}$]{{\includegraphics[width=\textwidth,height=3.5cm]{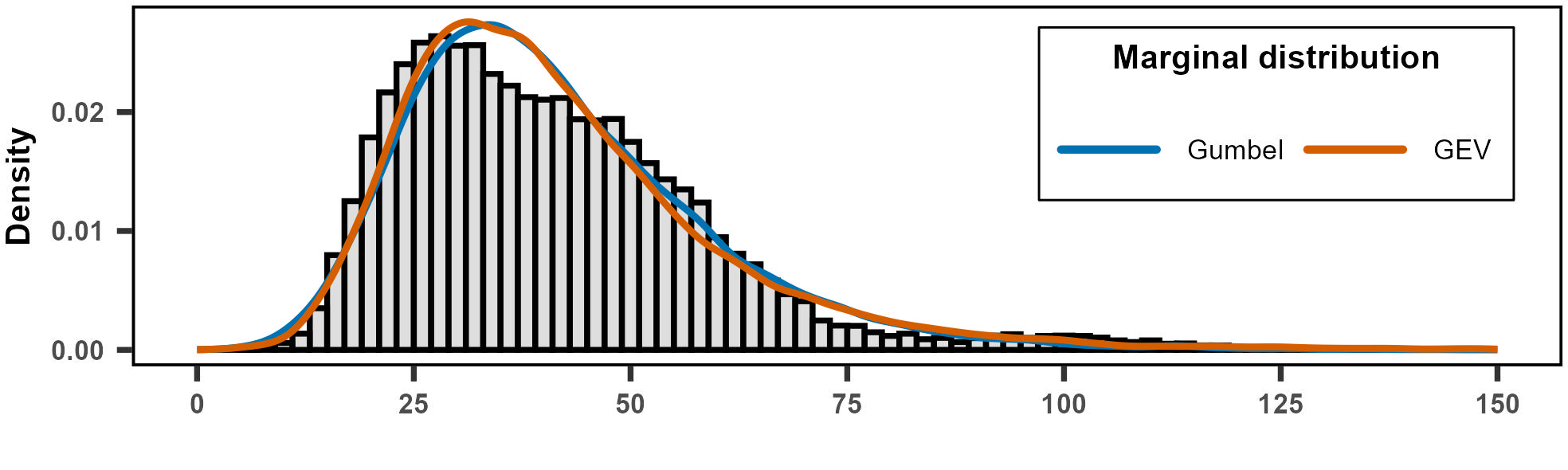}}}%
\hfill 
\subfigure[\fontsize{10}{15}\selectfont $\text{PM}_{2.5}$]{{\includegraphics[width=\textwidth,height=3.5cm]{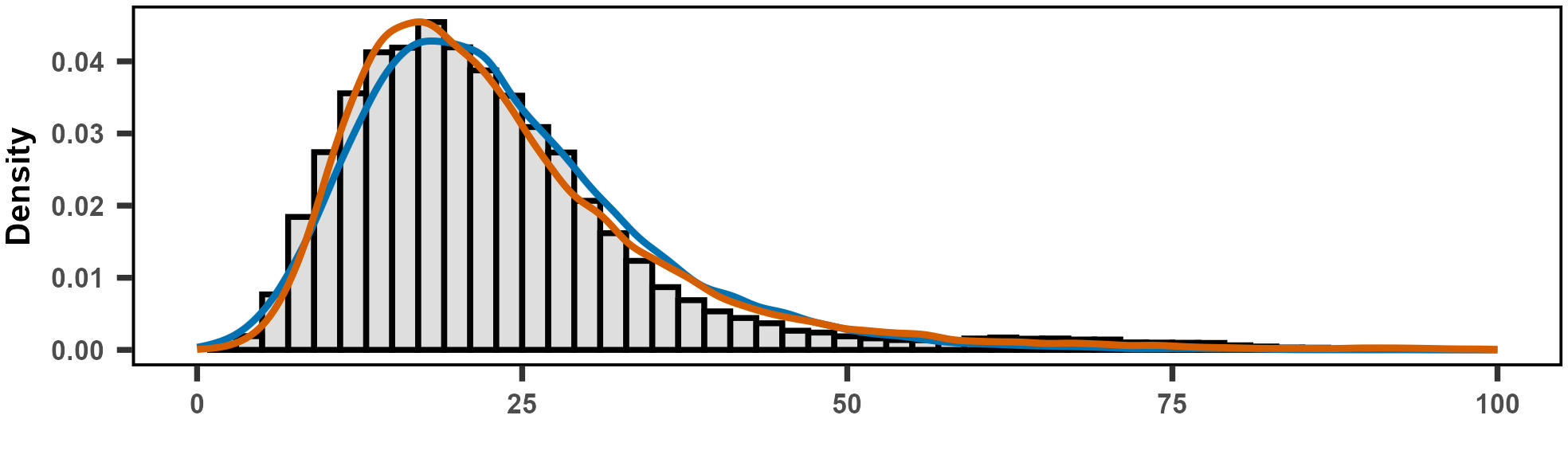}}}%
\hfill 
\subfigure[\fontsize{10}{15}\selectfont $\text{CO}$]{{\includegraphics[width=\textwidth,height=3.5cm]{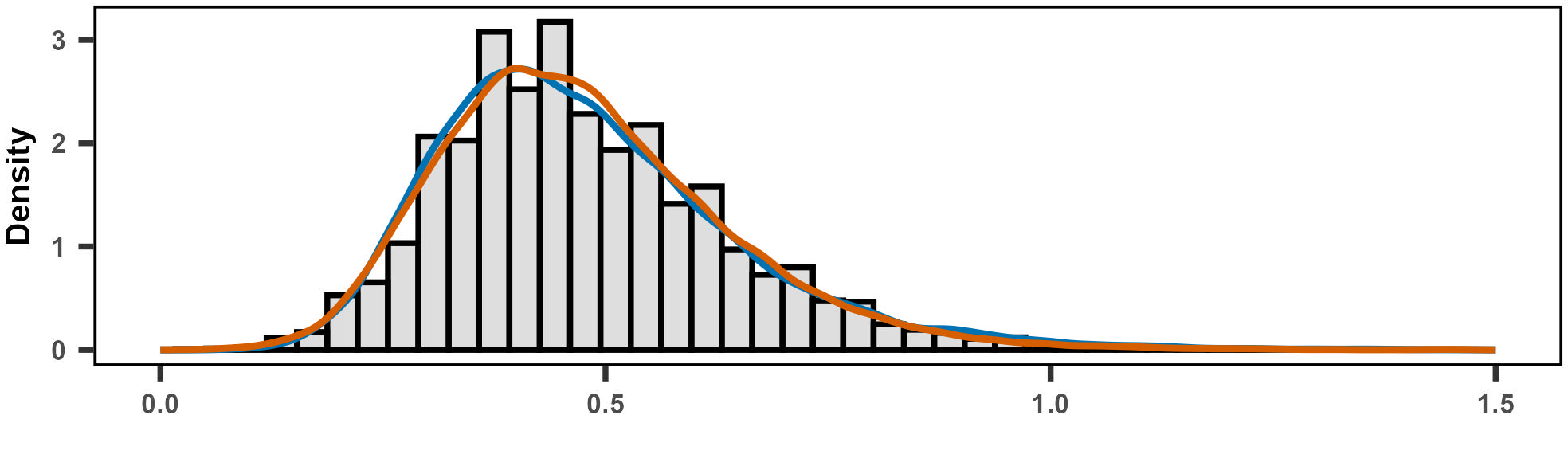}}}%
\caption{Histograms of the variables and transformation to uniform distribution.}\label{fig:fig 3}
\end{figure}

Figure 5 illustrates the spatio-temporal dependencies between $\text{PM}_{10}$ and two covariates, following the same approach as Figure 4. The dependence between $\text{PM}_{10}$ and $\text{PM}_{2.5}$ was examined for (a) dataset D1 and (b) dataset D2, while the dependence between $\text{PM}_{10}$ and $\text{CO}$ was analyzed for (c) dataset D1 and (d) dataset D2. The dependence between $\text{PM}_{10}$ and $\text{PM}_{2.5}$ was predominantly captured by the Joe copula, regardless of data type, marginal distribution, or temporal effects, with occasional selection of the Gumbel or T copula. This consistency suggested that tail dependence between these variables remained stable over time. In contrast, the dependence between $\text{PM}_{10}$ and $\text{CO}$ displayed a wider range of selections depending on specific conditions. In the dataset D1, the Joe copula was frequently selected when accounting for a one-time lag, whereas the Gumbel and T copulas were more commonly chosen for contemporaneous observations. In the dataset D2, the Joe copula remained dominant across different temporal structures, although the Frank, Gumbel, and T copulas were also selected depending on the assumed marginal distributions. Despite these patterns, copulas that captured tail dependence were consistently identified, emphasizing the presence of significant dependence in the tails.

\FloatBarrier 
\begin{figure}[h]
    \centering
    \begin{subfigure}
        \centering
        \begin{tikzpicture}
        \node[anchor=south west,inner sep=0] (image) at (0,0) {
            \includegraphics[width=\textwidth]{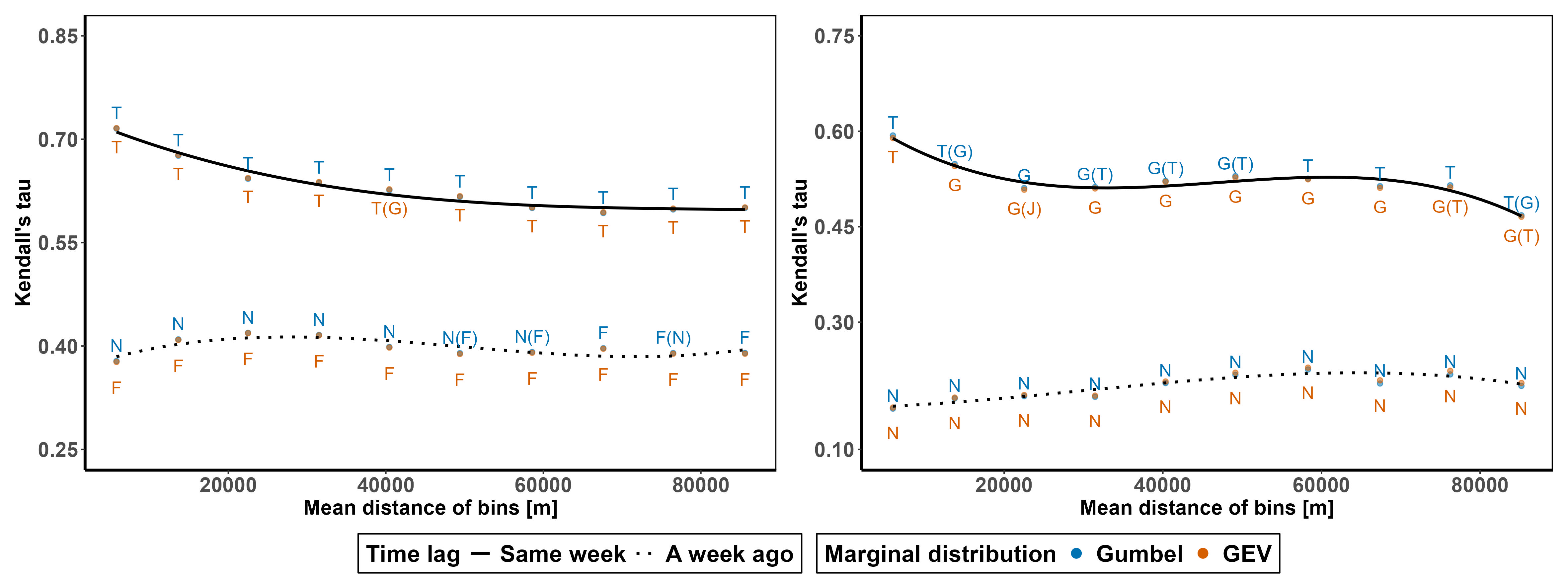}
        };
        \node[align=center] at (5, -0.3) {(a) The dataset D1};
        \node[align=center] at (13,-0.3){(b) The dataset D2};        
    \end{tikzpicture}
    \end{subfigure} 
     \caption{Estimated polynomial graph according to the assummed distribution. Letters denote the chosen copula family: Gaussian (N), Student\textquotesingle s (T), Clayton (C), Frank (F), Gumbel (G), Joe (J). For each spatial bin in the 10-fold CV, the most frequently selected copula is listed first, followed by the second most frequently selected copula in parentheses, if present.}\label{fig:fig 4} 
\end{figure}

\FloatBarrier 
\begin{figure}[h]
    \centering
    \begin{tikzpicture}
        \node[anchor=south west,inner sep=0] (image1) at (0,4) {
            \includegraphics[width=\textwidth]{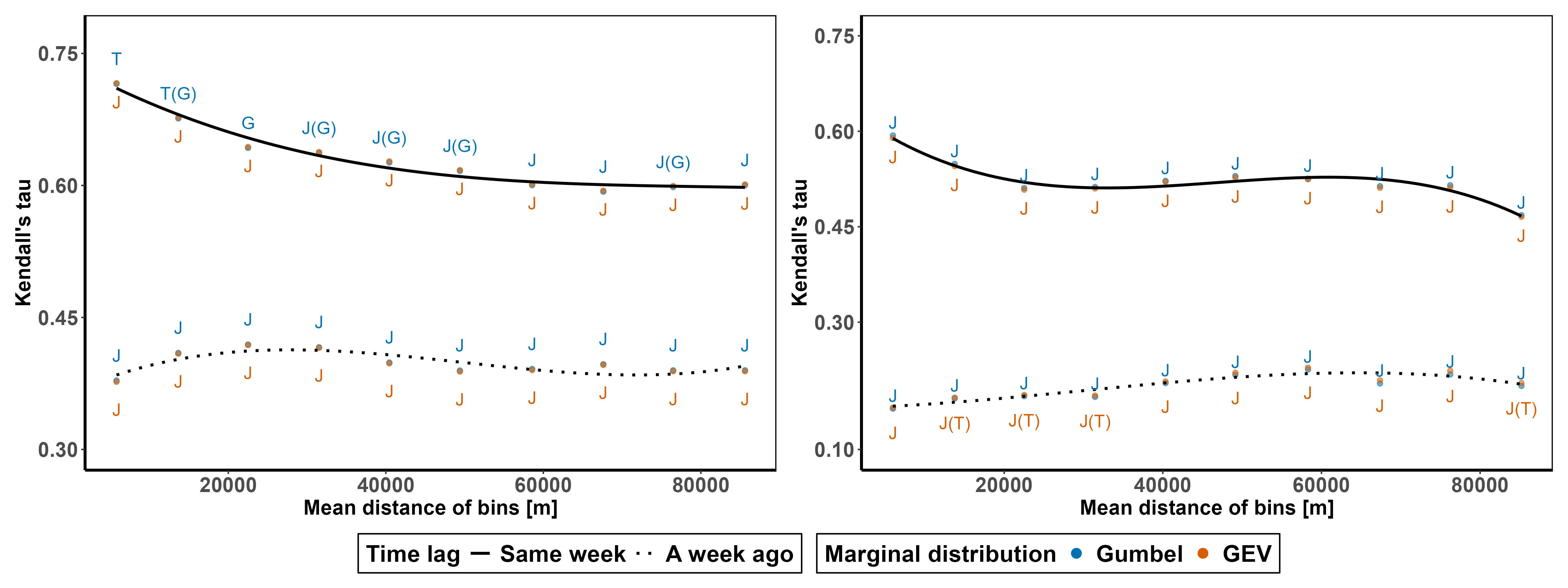}
        };
        \node[align=center] at (5,3.5) {(a) The dataset D1 for $\text{PM}_{2.5}$};
        \node[align=center] at (13,3.5) {(b) The dataset D2 for $\text{PM}_{2.5}$};
        \node[anchor=south west,inner sep=0] (image2) at (0,-3) {
            \includegraphics[width=\textwidth]{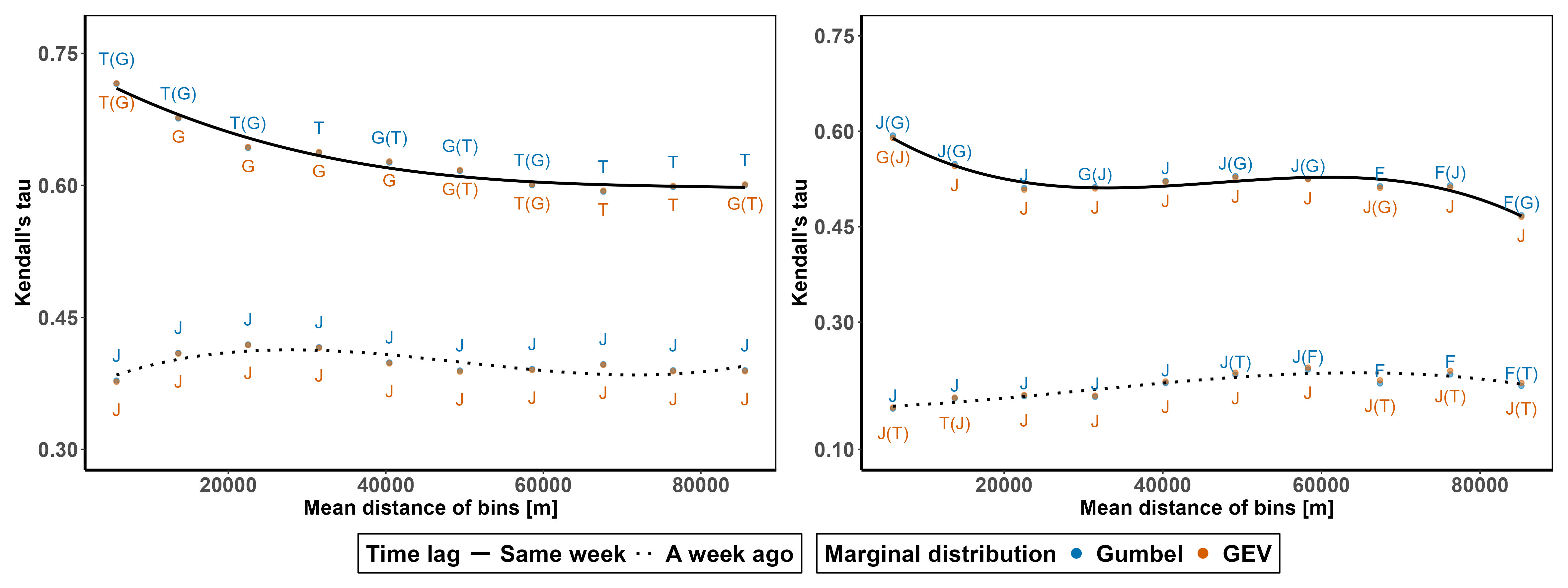}
        };
        \node[align=center] at (5,-3.5) {(c) The dataset D1 for $\text{CO}$};
        \node[align=center] at (13,-3.5) {(d) The dataset D2 for $\text{CO}$};
    \end{tikzpicture}

    \caption{Estimated polynomial graphs based on models including each covariate individually: $\text{PM}_{2.5}$ and $\text{CO}$. The copula representation follows that of Figure 4.}
    \label{fig:fig5}
\end{figure}

\subsection{Estimation of subsequent trees in the proposed model: conditional copula}
\label{subsec4.3}

The construction of the proposed model requires defining spatio-temporal neighborhood structures. In South Korea, $\text{PM}_{10}$ monitoring stations are unevenly distributed based on factors such as population density and geography. Considering the irregular spacing between stations, the spatial neighborhood for $\text{PM}_{10}$ was defined as the three nearest regions surrounding a central location. For covariates, only one neighboring region was included, as their influence was assumed to be weaker compared to $\text{PM}_{10}$. To incorporate temporal dependencies, observations from the same time point and a one-time lag were considered. As a result, each location had a total of ten spatio-temporal neighbors: six related to $\text{PM}_{10}$ and four associated with covariates. These variables were used to construct the subsequent trees of the proposed vine model, resulting in $\frac{10 \times 9}{2} = 45$ bivariate copulas. For each dataset, a total of $450$ bivariate copula estimates were obtained through 10-fold CV, where each fold contributed $45$ estimates.

Figure 6 presents the results of $450$ estimated bivariate copulas under different distributional assumptions for each dataset, highlighting the differences in copula selection across datasets. In the dataset D1, the T copula was most frequently selected, with 216 cases (54.67\%) under the Gumbel distribution and 209 cases (46.43\%) under the GEV distribution. The Clayton copula, including rotated variants, accounted for approximately 15.11\% of the selections under the Gumbel distribution assumption, while it was chosen for 22.89\% of the the selections under the GEV distribution assumption. Other copula families were selected less frequently. In the dataset D2, the Clayton copula dominated, with 116 cases (25.78\%) under the Gumbel distribution and 159 cases (35.33\%) under the GEV distribution, followed by the T copula as the second most common choice. These findings revealed the importance of considering a broad range of copula families, including rotated versions, to capture diverse dependency structures. Moreover, they illustrated the versatility of vine copulas in modeling complex and heterogeneous spatio-temporal relationships.

\subsection{Estimation of comparative models: traditional Bayesian spatio-temporal model, universial kriging, spatio-temporal Gaussian vine copula, and STCV}
\label{subsec4.4}

In this study, four benchmark models were selected for comparison. The first model was a traditional Bayesian spatio-temporal model. The parameters $\beta_{0}$, $\beta_{1}$, and $\beta_{2}$ were assigned normal priors with a mean of zero and a large variance, while the remaining parameters $\phi$, $\epsilon_{\gamma}^{2}$, $\epsilon_{\varepsilon}^{2}$, $\rho$, $\kappa$, $\epsilon_{u}^{2}$, and $\epsilon_{e}^{2}$ followed uniform priors over the range $[0, 10]$. For all data types, the posterior means were utilized as Bayesian estimates, and except for $\beta_{0}$ in the dataset D1 and $\beta_{0}$ and $\beta_{2}$ in the dataset D2, all other estimates were considered significant, as their 95\% credible intervals did not include zero. Detailed information was provided in Appendix Table B.1, where each Bayesian estimate and its corresponding 95\% credible interval were indicated. Moreover, the Gelman-Rubin diagnostic values were all below $1.05$, indicating a successful convergence of the MCMCs. Using these estimated parameters, the outcome $\hat{y}_{st}$ was predicted for unobserved spatio-temporal regions by utilizing the covariate values for these areas. However, some of these predictions for $\text{PM}_{10}$ concentrations yielded negative values, reflecting limitations in the predictive capacity of the model for certain regions.

\FloatBarrier 
\begin{figure}[h]
\centering
\includegraphics[width=\textwidth]{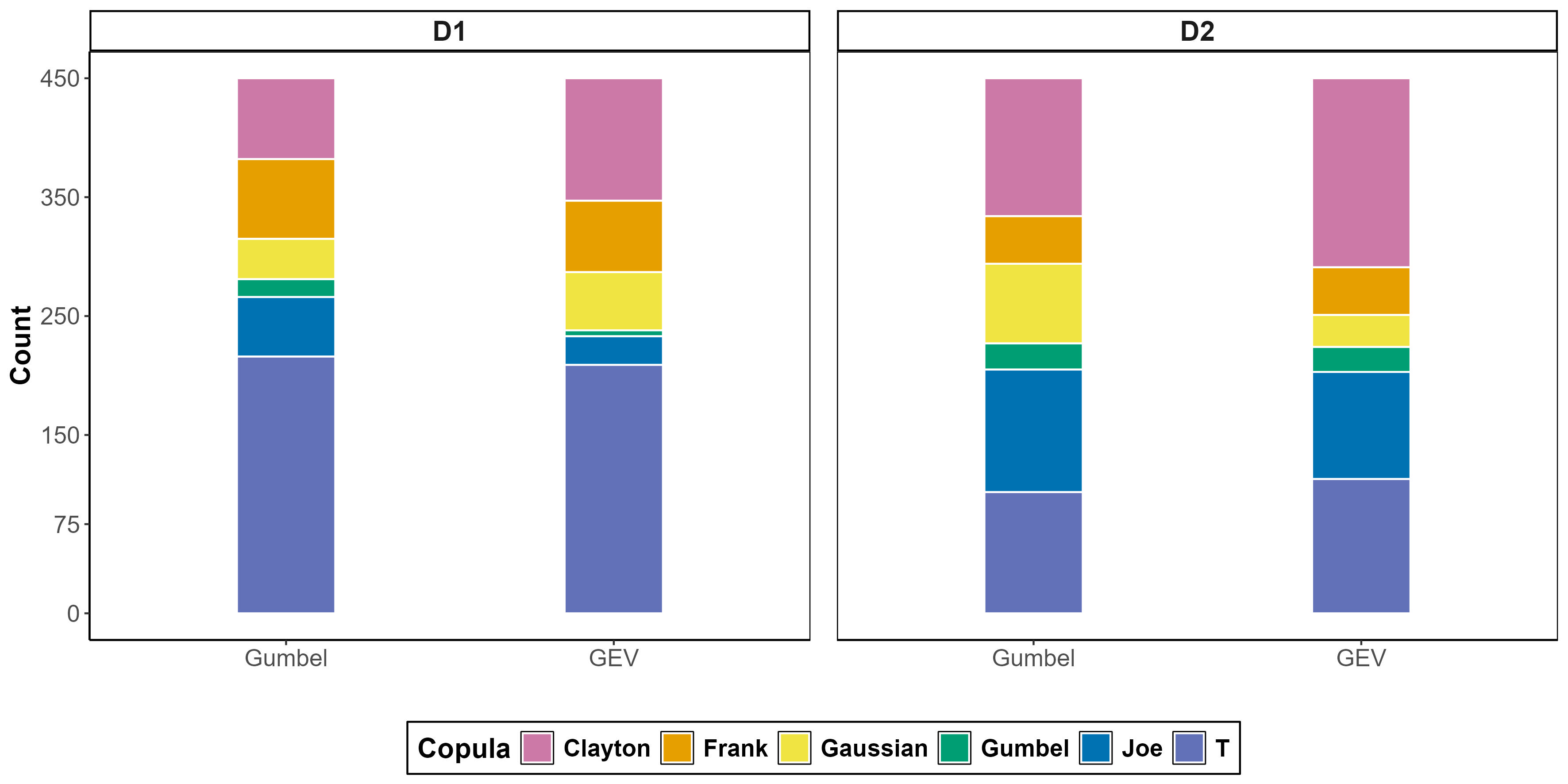} 
\caption{Results of the upper trees of the proposed model according to the assumed marginal distributions. Results assuming Gumbel distribution on the left and GEV distribution on the right.}\label{fig:fig 6}
\end{figure}

The second model was universal kriging, where variograms were computed for the same number of spatial bins as the correlogram. At each time point, the covariance structure was determined by selecting the best-fitting model—spherical, exponential, or Gaussian—based on the minimum sum of squared errors. The third model was a spatio-temporal Gaussian vine copula, which assumed that all dependencies could be described using Gaussian copulas. The final model was the STCV, with the same conditions of spatial bins, temporal effects, and the same spatio-temporal neighborhood for $\text{PM}_{10}$, as considered in this study.

\subsection{Prediction results}
\label{subsec4.5}

Table 2 presents a comparison of the predictive performance of the proposed model with four alternative methods based on MAE and RMSE. The performance in predicting extreme values defined as the top 5\% within each fold is presented in parentheses. The proposed model consistently outperformed kriging and STCV for both overall data and extreme value prediction across all datasets. The proposed model was generally more competitive than the Bayesian model, excluding extreme value prediction in the dataset D2. Notably, the higher RMSE values observed for the Bayesian model indicated greater variability compared to the proposed method. Although the spatio-temporal Gaussian vine copula model slightly outperformed the proposed approach in terms of overall prediction accuracy, the proposed approach yielded superior results for predicting extreme values. These results emphasized the advantages of the copula-based framework over traditional spatial prediction methods, such as Bayesian models or kriging, particularly in modeling and predicting extreme events.

\FloatBarrier 
\begin{table}[H]
\caption{MAE and RMSE of the four models for two datasets. The values in parentheses represent the performance metrics for predicting the top 5\% extreme values of each fold.}\label{table:table2}
\resizebox{\textwidth}{!}{  
\begin{tabular}{@{}cccccc@{}}
\toprule
\multirow{2}{*}{\textbf{Model}} & \multicolumn{1}{c}{\textbf{Marginal}} & \multicolumn{2}{c}{\textbf{Dataset D1}} & \multicolumn{2}{c}{\textbf{Dataset D2}}\\
\cmidrule(lr){3-4} \cmidrule(lr){5-6} &  & \textbf{MAE} & \textbf{RMSE} & \textbf{MAE} & \textbf{RMSE}\\
\midrule
\multirow{2}{*}{\textbf{Proposed model}} & \textbf{Gumbel} & 17.33 (44.63)& 22.38 (48.58) & 19.24 (52.41) &25.27 (54.62) \\
  & \textbf{GEV} & 18.35 (44.15) & 24.44 (48.33) &  18.12 (53.48) & 23.70 (55.08)\\
 \multirow{2}{*}{\textbf{STCV model}} & \textbf{Gumbel} & 18.56 (47.34)&24.15 (52.11)  & 22.24 (53.83)  & 29.42 (56.86) \\
  & \textbf{GEV} & 18.45 (47.45) & 24.00 (52.12) & 22.24 (53.07)  &  29.42 (55.61)\\
 \multirow{2}{*}{\textbf{Spatio-temporal Gaussian vine model}} & \textbf{Gumbel} & 15.02 (49.01)& 19.68 (52.02) & 17.55 (53.25) & 22.98 (54.70) \\
  & \textbf{GEV} & 14.91 (49.63) & 19.53 (52.55) &  17.23 (53.88)  &22.75 (55.17)\\
  \multicolumn{2}{c}{\textbf{Universal Kriging}} &  18.53 (49.00) & 28.16 (57.06) &21.67 (54.39) &28.98 (57.54)\\
 \multicolumn{2}{c}{\textbf{Bayesian spatio-temporal}} &  32.63 (70.13) & 39.99 (71.97) & 23.60 (32.65) & 26.80 (34.82) \\
\bottomrule
\end{tabular}
}
\end{table}

Table 3 summarizes the number and proportion of cases that fell within the 95\% prediction intervals across all spatial and temporal domains for each model, presented separately for the entire dataset and the subset of extreme values. For the dataset D1, the proposed model exhibited the highest coverage of observed values within the 95\% prediction intervals compared to other models, for both the full dataset and the extreme value subset. When assuming the Gumbel distribution, the proposed model captured 56.97\% of the observed values (11,883 cases), while under the GEV assumption, the coverage was 51.17\% of the observed values (10,674 cases). These values were relatively higher than those of competing models, and a similar trend was observed for the extreme value subset. Regardless of the marginal distribution assumption, the proposed model yielded prediction intervals that encompassed approximately 60.16\% and 52.38\% of the extreme values under the Gumbel and GEV distributions, respectively. In contrast, for the dataset D2, this superior performance was observed only when assuming the Gumbel distribution. When the GEV distribution was assumed, the proposed model still included more observed values within its prediction intervals compared to Bayesian or kriging models; however, some other copula-based models exhibited higher coverage. Even so, these results suggested that the proposed model provided well-calibrated prediction intervals and was more effective than other models in capturing both the overall data distribution and extreme values within the intervals.

\begin{table}[H]
\caption{Number of observed values that fell within the 95\% prediction intervals.
The results are shown for both the entire dataset (entire) and for the subset consisting of extreme values (extreme). Each cell presents the number of cases, with the percentage shown in parentheses.}
\label{table:table3}
\resizebox{\textwidth}{!}{  
\centering
\begin{tabular}{@{}cccccc@{}}
\toprule
\multirow{2}{*}{\textbf{Model}} & \multicolumn{1}{c}{\textbf{Marginal}} & \multicolumn{2}{c}{\textbf{Dataset D1}} & \multicolumn{2}{c}{\textbf{Dataset D2}}\\
\cmidrule(lr){3-4} \cmidrule(lr){5-6} 
&  & \textbf{Entire} & \textbf{Extreme} & \textbf{Entire} & \textbf{Extreme}\\
\midrule
\multirow{2}{*}{\textbf{Proposed model}} & \textbf{Gumbel} & 11883 (56.97) & 657 (60.16) & 1392 (56.86) & 75 (59.52)\\
& \textbf{GEV} & 10674 (51.17) & 572 (52.38) & 1358 (55.47) & 72 (57.14)\\
\multirow{2}{*}{\textbf{STCV model}} & \textbf{Gumbel} & 9738 (46.68) & 501 (45.88) & 1153 (47.10) & 53 (42.06)\\
& \textbf{GEV} & 9898 (47.45) & 521 (47.71) & 1432 (58.50) & 74 (58.73)\\
\multirow{2}{*}{\textbf{Spatio-temporal Gaussian vine model}} & \textbf{Gumbel} & 9545 (45.76) & 521 (47.71) & 1325 (54.13) & 71 (56.35)\\
& \textbf{GEV} & 9528 (45.68) & 519 (47.53) & 1311 (53.55) & 73 (57.94)\\
\multicolumn{2}{c}{\textbf{Universal Kriging}} & 8975 (43.03) & 457 (41.85) & 978 (39.95) & 43 (34.13)\\
\multicolumn{2}{c}{\textbf{Bayesian spatio-temporal}} & 862 (4.13) & 41 (3.75) & 284 (11.60) & 11 (8.73)\\
\bottomrule
\end{tabular}
}
\end{table}

We evaluated whether the proposed model outperformed alternative models even when considering a single covariate. Detailed results are provided in Appendix C for $\text{PM}_{2.5}$ and Appendix D for $\text{CO}$. Regardless of the covariate considered, the proposed model outperformed the three alternative methods, except for the spatio-temporal Gaussian vine copula, in both overall and extreme value prediction for the dataset D1. For the dataset D2, the proposed model consistently achieved higher predictive accuracy than both the Bayesian and kriging models, irrespective of the covariate. However, for extreme value prediction, the three copula-based models exhibited comparable performance. Across both datasets, the proposed model more frequently captured the observed values within the 95\% prediction intervals. Overall, these results suggested that copula-based models, including the proposed approach, effectively captured extreme values and achieved robust predictive performance across the entire dataset, even when using a single covariate.

\section{Conclusion and Discussion}
\label{sec5}

This study focused on spatio-temporal copula modeling of $\text{PM}_{10}$ concentrations in South Korea for 2019, aiming to predict the concentrations in unmeasured regions. To effectively capture the complex dependencies across time and space, as well as predict extreme values, the concept of spatio-temporal neighbors and copula modeling were employed. We proposed incorporating the spatio-temporal neighborhood relationships between $\text{PM}_{10}$ and two covariates, $\text{PM}_{2.5}$ and $\text{CO}$, for a better understanding of the dynamic relationships between dependent variable and covariates. In other words, we devised a methodology that incorporates the spatio-temporal neighborhood structures of $\text{PM}_{10}$ and the spatio-temporal neighborhood relationships between $\text{PM}_{10}$ and two covariates, $\text{PM}_{2.5}$ and $\text{CO}$, to model their dependencies. To simplify these complex and multidimensional dependencies, we employed vine copulas that decompose multivariate dependencies into bivariate structures. Given the left-skewed distributions of all three variables, Gumbel and GEV distributions were used as marginal distributions, transforming the data into uniform margins. Furthermore, analysis was performed using data from national monitoring stations in 2019, as well as a subset of data in SMA from January to April, which exhibited higher $\text{PM}_{10}$ concentrations. The proposed spatio-temporal vine copula model successfully captured complex and rich dependency structures, including symmetric and asymmetric patterns, as well as tail dependencies in both upper and lower extreme values. Model evaluation was performed through comparison with the Bayesian spatio-temporal model, universal kriging, and two copula-based models, namely the spatio-temporal Gaussian vine copula and the model proposed by \citet{graler2014developing}. The proposed model exhibited slightly higher MAE and RMSE values in overall prediction for certain datasets than other two copula-based models. However, it excelled in extreme value prediction by explicitly accounting for the spatio-temporal relationships between $\text{PM}_{10}$ and $\text{PM}_{2.5}$, as well as between $\text{PM}_{10}$ and $\text{CO}$, while effectively capturing strong correlations in the tail regions. Additionally, the proposed model outperformed both the Bayesian model and kriging in terms of predictive accuracy for both overall cases and extreme value scenarios.

This study is particularly noteworthy, as it addresses the absence of prior research in South Korea on modeling complex spatio-temporal dependencies in $\text{PM}_{10}$ data using copulas, a methodological approach not previously employed in this context. We proposed incorporating spatio-temporal relationships between the dependent variable and covariates, thereby providing a comprehensive analysis of the copula-based dependency structure. This approach captured both spatial and temporal dynamics with covariates, which were not considered in previous studies, thereby enhancing the modeling of complex dependencies and improving the accuracy and interpretability of the multivariate relationships of $\text{PM}_{10}$. Moreover, the proposed method facilitated seamless extension to a multivariate framework by utilizing vine copulas, effectively modeling the complex interdependencies inherent in high-dimensional spatio-temporal data. Kriging assumes a Gaussian distribution, limiting its ability to capture complex dependencies. In contrast, our model accounted for non-Gaussian and asymmetric structures, providing more realistic estimates. The proposed copula-based approach consistently produced reliable estimates, avoiding the potentially unreasonable predictions that may arise in Bayesian models. Notably, the model exceled in capturing tail dependencies for extreme values by utilizing copulas capable of representing such relationships, in contrast to Gaussian copulas, which lack this property. The model\textquotesingle s exceptional predictive performance in the tail region enhanced the credibility of spatio-temporal modeling and underscored its potential to improve predictive validity in environmental studies and public health assessments. Additionally, the vine copula framework demonstrated robustness in sensitivity analysis by maintaining consistent results across varying covariate orders, further validating its adaptability and reliability. Moreover, the use of vine copulas significantly enhanced computational efficiency compared to Bayesian modeling, and the proposed model exhibited excellent performance even when a single covariate was considered.

We identified several limitations in our study. First, the selection of covariate types, the number of time lags, the number of spatial bins, and the number of nearest neighbors were based more on empirical judgment than on systematic methodologies. Employing a more structured strategy for determining these parameters could enhance not only the generalizability of the model results but also their overall effectiveness. Second, while numerous factors can influence $\text{PM}_{10}$ concentrations, obtaining reliable data for such factors remains challenging. For instance, meteorological data are prone to measurement errors due to various reasons, and the spatial distribution of meteorological stations in South Korea is limited. Furthermore, meteorological and $\text{PM}_{10}$ monitoring stations are often located in different areas, making it difficult to integrate spatial information from both sources. These constraints pose major challenges to collecting appropriate covariate data. Third, the correlation between the spatio-temporal neighbors of $\text{PM}_{10}$ and the covariates was assumed to be independent in this study. However, this assumption may compromise accuracy and generalizability by overlooking the complexity of their relationships.

To address these limitations, future research should establish systematic strategies for parameter selection that effectively leverage spatio-temporal neighbor information. Additionally, incorporating other air pollutants known to affect $\text{PM}_{10}$ concentrations, such as $\text{O}_{3}$, $\text{NO}_{2}$, and meteorological variables like precipitation and average humidity as covariates would be beneficial. This would help mitigate unrealistic outcomes in $\text{PM}_{10}$ concentration forecasts. The model\textquotesingle s strong performance in predicting extreme values suggests its potential for application to other extreme value datasets to further evaluate its efficacy.

\section*{Funding}
\label{Funding}
This study was supported by the National Research Foundation of Korea (NRF) grant funded by the Korea government (MSIT) (No. NRF-2021R1A2C1010595).

\section*{Data Availability}
\label{Data Availability}
Data for this study can be found at https://github.com/soyunjeon-stat/STCV.

\appendix
\section*{Appendix A: Supplementary Information}
\label{app1}
\setcounter{equation}{0}  
\renewcommand{\theequation}{A.\arabic{equation}}  
\textbf{A1. Gumbel CDF with parameters location $a$ and scale $b$ is, for all real $z$,}
\begin{equation}
    F(x)=\exp \{-\exp[-\frac{x-a}{b}]\}, \text{where all real}\  a \ \text{and} \  b>0 
\end{equation}

\

\textbf{A2. GEV CDF with parameters location $a$, scale $b$, and shape $s$ is, for $1+\frac{s(x-a)}{b}>0$,} 
\begin{equation}
    F(x)=\exp[-\{1+\frac{s(x-a)}{b}\}^{-1/s}], \text{where all real} \ a,s \ \text{and} \  b>0 
\end{equation}

\FloatBarrier

\section*{Appendix B: The estimated result of the Bayesian spatio-temporal model}
\label{app2}
\setcounter{table}{0}
\renewcommand{\thetable}{B.\arabic{table}}
\renewcommand{\arraystretch}{1.3}  
\setlength{\tabcolsep}{5.5pt}       

\begin{center}
\captionsetup{justification=raggedright, singlelinecheck=false,font={sf}}
\captionof{table}{The estimated results of the Bayesian spatio-temporal model for two datasets D1 and D2.}
\label{table:app2}
\begin{adjustbox}{max width=\textwidth}
{\fontfamily{lmss}\selectfont  
\begin{tabular}{@{} c ccc ccc @{}}
\toprule
\multicolumn{1}{c}{\textbf{Parameter}} & \multicolumn{3}{c}{\textbf{Dataset D1}} & \multicolumn{3}{c}{\textbf{Dataset D2}} \\
\cmidrule(lr){2-4} \cmidrule(lr){5-7}
\multicolumn{1}{c}{} & \textbf{Mean} & \textbf{2.5\%} & \textbf{97.5\%} & \textbf{Mean} & \textbf{2.5\%} & \textbf{97.5\%} \\
\midrule
$\bm{\beta_{0}}$ &  \makebox[8mm][r]{0.525}  & \makebox[8mm][r]{-1.402}  & \makebox[8mm][r]{2.452}  & \makebox[8mm][r]{0.364}  & \makebox[8mm][r]{-1.580}  & \makebox[8mm][r]{2.318}  \\
$\bm{\beta_{1}}$ &  \makebox[8mm][r]{0.336}  & \makebox[8mm][r]{0.316}  & \makebox[8mm][r]{0.357}  & \makebox[8mm][r]{1.052}  & \makebox[8mm][r]{1.019}  & \makebox[8mm][r]{1.086}  \\
$\bm{\beta_{2}}$ &  \makebox[8mm][r]{2.556}  & \makebox[8mm][r]{1.313}  & \makebox[8mm][r]{3.799}  & \makebox[8mm][r]{1.806}  & \makebox[8mm][r]{-0.457}  & \makebox[8mm][r]{4.069}  \\
$\bm{\phi}$ &  \makebox[8mm][r]{0.952}  & \makebox[8mm][r]{0.880}  & \makebox[8mm][r]{0.997}  & \makebox[8mm][r]{0.904}  & \makebox[8mm][r]{0.739}  & \makebox[8mm][r]{0.996}  \\
$\bm{\epsilon_{\gamma}^{2}}$ &  \makebox[8mm][r]{7.870}  & \makebox[8mm][r]{5.356}  & \makebox[8mm][r]{9.840}  & \makebox[8mm][r]{3.864}  & \makebox[8mm][r]{0.200}  & \makebox[8mm][r]{8.991}  \\
$\bm{\epsilon_{u}^{2}}$ &  \makebox[8mm][r]{0.364}  & \makebox[8mm][r]{0.021}  & \makebox[8mm][r]{1.102}  & \makebox[8mm][r]{2.203}  & \makebox[8mm][r]{0.200}  & \makebox[8mm][r]{3.771}  \\
$\bm{\epsilon_{\varepsilon}^{2}}$ &  \makebox[8mm][r]{8.786}  & \makebox[8mm][r]{6.679}  & \makebox[8mm][r]{9.954}  & \makebox[8mm][r]{8.346}  & \makebox[8mm][r]{5.480}  & \makebox[8mm][r]{9.928}  \\
$\bm{\epsilon_{e}^{2}}$ &  \makebox[8mm][r]{8.923}  & \makebox[8mm][r]{8.833}  & \makebox[8mm][r]{9.015}  & \makebox[8mm][r]{5.408}  & \makebox[8mm][r]{5.246}  & \makebox[8mm][r]{5.577}  \\
$\bm{\kappa}$ &  \makebox[8mm][r]{0.053}  & \makebox[8mm][r]{0.050}  & \makebox[8mm][r]{0.062}  & \makebox[8mm][r]{0.489}  & \makebox[8mm][r]{0.069}  & \makebox[8mm][r]{0.927}  \\
$\bm{\rho}$ &  \makebox[8mm][r]{159.410}  & \makebox[8mm][r]{101.460}  & \makebox[8mm][r]{282.351}  & \makebox[8mm][r]{199.925}  & \makebox[8mm][r]{105.000}  & \makebox[8mm][r]{294.890}  \\
\bottomrule
\end{tabular}
}
\end{adjustbox}
\end{center}

\FloatBarrier
\clearpage
\section*{Appendix C: Prediction results of the five models using only covariate $\text{PM}_{2.5}$}
\label{app3}
\setcounter{table}{0}  
\renewcommand{\thetable}{C.\arabic{table}}  
\setlength{\tabcolsep}{5pt} 
\renewcommand{\arraystretch}{1.5} 
\FloatBarrier
\begin{center}
\captionsetup{justification=raggedright, singlelinecheck=false,font={sf}}
\captionof{table}{MAE and RMSE of the four models for two datasets using only covariate $\text{PM}_{2.5}$. The values in parentheses represent the performance metrics for predicting the top 5\% extreme values of each fold.}
\label{table:app3}
\begin{adjustbox}{max width=\textwidth}
{\fontfamily{lmss}\selectfont  
\begin{tabular}{@{}cccccc@{}}
\toprule
\multirow{2}{*}{\textbf{Model}} & \multicolumn{1}{c}{\textbf{Marginal}} & \multicolumn{2}{c}{\textbf{Dataset D1}} & \multicolumn{2}{c}{\textbf{Dataset D2}}\\
\cmidrule(lr){3-4} \cmidrule(lr){5-6} &  & \textbf{MAE} & \textbf{RMSE} & \textbf{MAE} & \textbf{RMSE}\\
\midrule
\multirow{2}{*}{\textbf{Proposed model}} & \textbf{Gumbel} & 17.07 (45.56)& 22.22 (48.48) & 19.57 (52.36) & 25.91 (54.74)    \\
  & \textbf{GEV} & 18.36 (44.90) & 24.62 (49.12) & 17.89 (53.75) & 23.49 (55.36)   \\
 \multirow{2}{*}{\textbf{STCV model}} & \textbf{Gumbel} & 18.85 (47.52)&24.71 (52.28)  & 22.28 (53.47)  &29.70 (56.67)  \\
  & \textbf{GEV} & 18.72 (47.59) & 24.45 (52.33) &  21.50 (51.63) & 27.77 (54.88) \\
 \multirow{2}{*}{\textbf{Spatio-temporal Gaussian vine model}} & \textbf{Gumbel} & 15.04 (50.86)& 19.82 (53.79) & 17.75 (52.86) &23.01 (54.21)  \\
  & \textbf{GEV} & 14.86 (51.73) & 19.68 (54.49) &  17.24 (54.03)  & 22.74 (55.28)\\
  \multicolumn{2}{c}{\textbf{Universal Kriging}} &  18.55 (49.13) & 28.60 (59.49) & 21.65 (54.37)& 28.96 (57.55)\\
 \multicolumn{2}{c}{\textbf{Bayesian spatio-temporal}} &  33.24 (71.45) & 40.27 (73.13) & 916.69 (835.23) &3494.98 (3312.43)  \\
\bottomrule
\end{tabular}
}
\end{adjustbox}
\end{center}

\begin{center}
\captionsetup{justification=raggedright, singlelinecheck=false,font={sf}}
\captionof{table}{Number of observed values that fell within the 95\% prediction interval when using only covariate $\text{PM}_{2.5}$.
The results are shown for both the entire dataset (entire) and for the subset consisting of extreme values (extreme). Each cell presents the number of cases, with the percentage shown in parentheses.}
\label{table:app4}  
\begin{adjustbox}{max width=\textwidth}
{\fontfamily{lmss}\selectfont  
\begin{tabular}{@{}cccccc@{}}
\toprule
\multirow{2}{*}{\textbf{Model}} & \multicolumn{1}{c}{\textbf{Marginal}} & \multicolumn{2}{c}{\textbf{Dataset D1}} & \multicolumn{2}{c}{\textbf{Dataset D2}}\\
\cmidrule(lr){3-4} \cmidrule(lr){5-6} 
&  & \textbf{Entire} & \textbf{Extreme} & \textbf{Entire} & \textbf{Extreme}\\
\midrule
\multirow{2}{*}{\textbf{Proposed model}} & \textbf{Gumbel} & 12101 (58.01) & 661 (60.53) & 1432 (58.50) & 79 (62.70)\\
& \textbf{GEV} & 10869 (52.11) & 577 (52.84) & 1406 (57.43) & 78 (61.90)\\
\multirow{2}{*}{\textbf{STCV model}} & \textbf{Gumbel} & 9651 (46.27) & 501 (45.88) & 1187 (48.49) & 57 (45.24)\\
& \textbf{GEV} & 983 (47.14) & 506 (46.34) &  1482 (60.54)& 74 (58.73)\\
\multirow{2}{*}{\textbf{Spatio-temporal Gaussian vine model}} & \textbf{Gumbel} & 8507 (40.78) & 447 (40.93) &  1200 (49.02) & 61 (48.41)\\
& \textbf{GEV} & 8590 (41.18) &  449 (41.12) & 1229 (50.20) & 63 (50.00)\\
\multicolumn{2}{c}{\textbf{Universal Kriging}} & 8889 (42.61) & 449 (41.12) & 975 (39.83) & 42 (33.33)\\
\multicolumn{2}{c}{\textbf{Bayesian spatio-temporal}} & 684 (3.28) & 29 (2.66) & 201 (8.21) &8 (6.35) \\
\bottomrule
\end{tabular}
}
\end{adjustbox}
\end{center}

\FloatBarrier
\clearpage
\section*{Appendix D: Prediction results of the five models using only covariate $\text{CO}$}
\label{app4}
\setcounter{table}{0}  
\renewcommand{\thetable}{D.\arabic{table}}  
\setlength{\tabcolsep}{5pt} 
\renewcommand{\arraystretch}{1.5} 

\FloatBarrier 
\begin{center}
\captionsetup{justification=raggedright, singlelinecheck=false,font={sf}}
\captionof{table}{MAE and RMSE of the four models for two datasets using only covariate $\text{CO}$. The values in parentheses represent the performance metrics for predicting the top 5\% extreme values of each fold.}
\label{table:app5}
\begin{adjustbox}{max width=\textwidth}
{\fontfamily{lmss}\selectfont  
\begin{tabular}{@{}cccccc@{}}
\toprule
\multirow{2}{*}{\textbf{Model}} & \multicolumn{1}{c}{\textbf{Marginal}} & \multicolumn{2}{c}{\textbf{Dataset D1}} & \multicolumn{2}{c}{\textbf{Dataset D2}}\\
\cmidrule(lr){3-4} \cmidrule(lr){5-6} &  & \textbf{MAE} & \textbf{RMSE} & \textbf{MAE} & \textbf{RMSE}\\
\midrule
\multirow{2}{*}{\textbf{Proposed model}} & \textbf{Gumbel} & 16.34 (45.84)& 21.03 (49.43) &  18.08 (53.19) & 23.65 (54.88)    \\
  & \textbf{GEV} & 17.56 (43.79) & 22.77 (47.95) & 17.58 (54.14) & 23.08 (55.40)   \\
 \multirow{2}{*}{\textbf{STCV model}} & \textbf{Gumbel} & 18.51 (47.72)&24.18 (52.09)&  22.44 (52.92) & 29.20 (55.92) \\
  & \textbf{GEV} & 18.37 (47.72) & 24.00 (52.27) &  21.94 (52.40) &28.47 (55.39)  \\
 \multirow{2}{*}{\textbf{Spatio-temporal Gaussian vine model}} & \textbf{Gumbel} & 14.91 (50.61)& 19.60 (53.44) & 17.74 (52.57) &22.98 (53.88)   \\
  & \textbf{GEV} & 14.83 (51.43) & 19.62 (54.16) &  17.43 (53.69)  &22.83 (54.89)   \\
  \multicolumn{2}{c}{\textbf{Universal Kriging}} &  19.29 (48.20) & 47.48 (52.50) &21.24 (54.95) &28.44 (57.87) \\
 \multicolumn{2}{c}{\textbf{Bayesian spatio-temporal}} &  36.34 (79.53) & 43.10 (81.40) & 1075.18 (1010.91) &4038.65 (3819.85)  \\
\bottomrule
\end{tabular}
}
\end{adjustbox}
\end{center}

\begin{center}
\captionsetup{justification=raggedright, singlelinecheck=false,font={sf}}
\captionof{table}{Number of observed values that fell within the 95\% prediction interval when using only covariate $\text{CO}$. The results are shown for both the entire dataset (entire) and for the subset consisting of extreme values (extreme). Each cell presents the number of cases, with the percentage shown in parentheses.}
\label{table:app6}
\begin{adjustbox}{max width=\textwidth}
{\fontfamily{lmss}\selectfont  
\begin{tabular}{@{}cccccc@{}}
\toprule
\multirow{2}{*}{\textbf{Model}} & \multicolumn{1}{c}{\textbf{Marginal}} & \multicolumn{2}{c}{\textbf{Dataset D1}} & \multicolumn{2}{c}{\textbf{Dataset D2}}\\
\cmidrule(lr){3-4} \cmidrule(lr){5-6} 
&  & \textbf{Entire} & \textbf{Extreme} & \textbf{Entire} & \textbf{Extreme}\\
\midrule
\multirow{2}{*}{\textbf{Proposed model}} & \textbf{Gumbel} &  12272 (58.83) & 687 (62.91) &  1471 (60.09) & 76 (60.32)\\
& \textbf{GEV} & 12756 (61.15) &  669 (61.26)&   1449 (59.19)  &77 (61.11) \\
\multirow{2}{*}{\textbf{STCV model}} & \textbf{Gumbel} & 10512 (50.40)  & 557 (51.01) & 1324 (54.08) &68 (53.97) \\
& \textbf{GEV} &  10929 (52.40) & 588 (53.85) &  1390 (56.78) & 71 (56.35)\\
\multirow{2}{*}{\textbf{Spatio-temporal Gaussian vine model}} & \textbf{Gumbel} & 8417 (40.35)  &446 (40.84)  &  1239 (50.61) & 57 (45.24)\\
& \textbf{GEV} & 8599 (41.22) &  459 (42.03) & 1225 (50.04) & 60 (47.62)\\
\multicolumn{2}{c}{\textbf{Universal Kriging}} & 12978 (62.22)  & 668 (61.17) & 1239 (50.61) & 55 (43.65) \\
\multicolumn{2}{c}{\textbf{Bayesian spatio-temporal}} & 524 (2.51)  &  28 (2.56)& 85 (3.47) &4 (3.17) \\
\bottomrule
\end{tabular}
}
\end{adjustbox}
\end{center}




\cleardoublepage

\bibliographystyle{apalike}
\bibliography{reference}

\end{document}